\documentclass[format=acmsmall, review=false, screen=true]{acmart}

\usepackage{booktabs} 

\usepackage{enumitem}
\usepackage{amssymb,amsfonts}
\usepackage{epsfig}
\usepackage[caption=false]{subfig}
\usepackage{url}
\usepackage{dsfont}

\begin{document}
\title{
End-to-End Entity Resolution for Big Data: A Survey 
}

\author{Vassilis Christophides}
\orcid{1234-5678-9012-3456}
\affiliation{%
  \institution{ENSEA, ETIS Lab}
  \country{France}
}
\email{vassilis.christophides@inria.fr}

\author{Vasilis Efthymiou}
\affiliation{%
  \institution{IBM Research}
  \country{USA}
}
\email{vasilis.efthymiou@ibm.com}

\author{Themis Palpanas}
\affiliation{%
  \institution{University of Paris \& French University Institute (IUF)}
  \country{France}
}
\email{themis@mi.parisdescartes.fr}

\author{George Papadakis}
\affiliation{%
  \institution{National and Kapodistrian University of Athens}
  \country{Greece}
}
\email{gpapadis@di.uoa.gr}

\author{Kostas Stefanidis}
\affiliation{%
  \institution{Tampere University}
  \country{Finland}
}
\email{konstantinos.stefanidis@tuni.fi}

\renewcommand{\shortauthors}{V. Christophides et al.}

\begin{abstract}
One of the most critical tasks for improving data quality and increasing the reliability of data analytics is \emph{Entity Resolution} (ER), which aims to identify different descriptions that refer to the same real-world entity. Despite several decades of research, ER remains a challenging problem. In this survey, we highlight the novel aspects of resolving Big Data entities when we should satisfy more than one of the Big Data characteristics simultaneously (i.e., Volume and Velocity with Variety). We present the basic concepts, processing steps and execution strategies that have been proposed by database, semantic Web and machine learning communities in order to cope with the loose \emph{structuredness}, extreme \emph{diversity}, high \emph{speed} and large \emph{scale} of entity descriptions used by real-world  applications. We provide an end-to-end view of ER workflows~for~Big Data, critically review the pros and cons of existing methods, and conclude with the main open research~directions.

\end{abstract}

\maketitle

\section{Introduction}

In the Big Data era, business, government and scientific organizations increasingly rely on massive amounts of data collected from both internal (e.g., CRM, ERP) and external data sources (e.g., the Web). Even when data integrated from multiple sources refer to the same real-world entities, they usually exhibit several quality issues such as \emph{incompleteness} (i.e., partial data), \emph{redundancy} (i.e., overlapping data), \emph{inconsistency} (i.e., conflicting data) or simply \emph{incorrectness} (i.e., data errors). A typical task for improving various aspects of data quality is \emph{Entity Resolution} (ER). 

\begin{figure*}
\begin{center}
\includegraphics[width=1.0\textwidth]{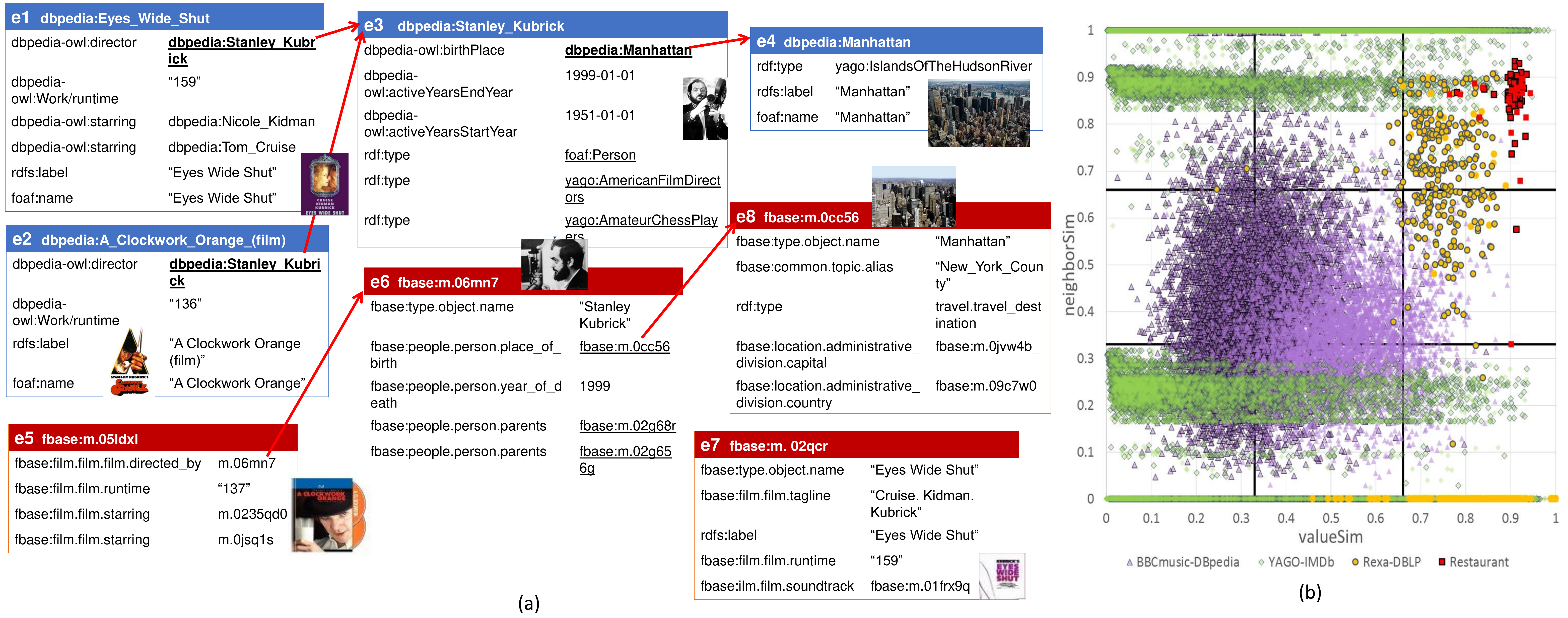}
\vspace{-18pt}
\caption{(a) Movies, directors and locations from DBpedia (blue) and Freebase (red), where $e_1$, $e_2$, $e_3$ and $e_4$ match with $e_7$, $e_5$, $e_6$ and $e_8$, resp. (b) Value and neighbor similarity distribution of matches in four datasets.
}
\label{fig:LOD++example}
\vspace{-14pt}
\end{center}
\end{figure*}

ER aims to identify different descriptions that refer to the same real-world entity appearing either within or across data sources, when unique entity identifiers are not available. Typically, ER aims to match structured descriptions (i.e., records) stored in the same (a.k.a., \emph{deduplication}), or two different (a.k.a., \emph{record linkage}) relational tables. In the Big Data era, other scenarios are also considered, such as matching semi-structured descriptions across RDF knowledge bases (KB) or XML-files (a.k.a., \emph{link discovery} or \emph{reference reconciliation}). Figure~\ref{fig:LOD++example}(a) illustrates descriptions of the same movies, directors and places from two popular KBs: DBpedia (blue) and Freebase (red). Each entity description is depicted in a tabular format, where the header row is the URI of the description and the remaining rows are the attribute (left) - value (right) pairs of the description.

ER aims to classify pairs of descriptions that are assumed to correspond to the same (vs. different) entity into \emph{matches} (vs. \emph{non-matches}). An ER process usually encompasses several tasks, including \emph{Indexing} (a.k.a., \emph{Blocking}), which reduces the number of candidate descriptions to be compared in detail, and \emph{Matching}, which assesses the similarity of pairs of candidate descriptions using a set of functions. Several ER frameworks and algorithms for these tasks have been proposed during the last three decades in different research communities. In this survey, we present the latest developments in ER, explaining how the Big Data characteristics call for novel ER frameworks that relax a number of assumptions underlying several methods and techniques proposed in the context of the database~\cite{DBLP:books/daglib/0030287,Dorneles2011,ElmagarmidIV07,DBLP:journals/pvldb/KopckeTR10,DBLP:series/synthesis/2010Naumann}, machine learning~\cite{Getoor:2012:ERT:2367502.2367564} and semantic Web communities~\cite{DBLP:journals/semweb/NentwigHNR17}. 

Our work is inspired by the Linked Open Data (LOD)  initiative~\cite{DBLP:series/synthesis/2015Christophides}, which covers only a small fragment of the Web today, but is representative of the challenges raised by Big Data to core ER tasks: $(a)$ how descriptions can be effectively compared for similarity, and $(b)$ how resolution algorithms can efficiently filter the number of candidate description pairs that need to be compared. 

\vspace{3pt}
\noindent
\textbf{Big Data Characteristics.} Entity descriptions published as LOD exhibit the  4 “V”s \cite{DBLP:series/synthesis/2015Dong} that challenge existing individual ER algorithms, but also entire ER workflows:

\begin{itemize}
\item \emph{Volume.} The content of each data source never ceases to increase and so does the \emph{number of data sources}, even for a single domain. For example, the LOD cloud currently contains more than 1,400 datasets from various sources (this is an x100 growth since its first edition) in 10 domains with $>$200B triples (i.e., $<subject, predicate, object>$) describing more than 60M entities of different types\footnote{\label{fn:lod}\url{https://lod-cloud.net}}; the life-science domain alone accounts for $>$350 datasets. 
\item \emph{Variety.} Data sources are extremely heterogeneous, even in the same domain, regarding both how they structure their data and how they describe the same real-world entity. In fact, they exhibit \emph{considerable diversity} even for substantially similar entities. For example, there are $\sim$700 vocabularies in the LOD cloud, but only $\sim$100 of them are shared by more than~one~KB\footnote{\label{fn:lod-vocabularies}\url{https://lov.linkeddata.es/dataset/lov}}.
\item \emph{Velocity.} As a direct consequence of the rate at which data is being collected and continuously made available, many of the data sources are \emph{very dynamic}. For example, LOD data are rarely static, with recent studies reporting that 23\% of the datasets exhibit infrequent changes, while 8\% are highly dynamic in terms of triples additions and  deletions\footnote{\label{fn:lod-dynamics}\url{http://km.aifb.kit.edu/projects/dyldo}}. 
\item \emph{Veracity.} Data sources are of \emph{widely differing quality}, with significant differences in the coverage, accuracy and timeliness of data provided. Even in the same domain, various forms of inconsistencies and errors in entity descriptions may arise, due to the limitations of the automatic extraction techniques, or of the crowd-sourced contributions. A recent empirical study \cite{DBLP:journals/semweb/Debattista0AC18} shows that there are several LOD quality problems, as their conformance with a number of best practices and guidelines is still open. For example, in Figure~\ref{fig:LOD++example}(a), the descriptions of ``A Clockwork Orange" from DBpedia ($e_2$) and Freebase ($e_5$)  differ~in~their~runtime.
\end{itemize}

\vspace{3pt}
\noindent
\textbf{Big Data Entity Resolution.} 
Individual characteristics of Big Data have been the focus of previous research work in ER. For example, there is a continuous concern for improving the \emph{scalability} of ER techniques over increasing \textit{Volumes} of entities using massively parallel implementations~\cite{DBLP:journals/ojbd/ChenSS18}. Moreover, uncertain entity descriptions due to high \textit{Veracity} have been resolved using approximate matching~\cite{Dorneles2011,DBLP:journals/pvldb/Gal14}. However, traditional deduplication techniques~\cite{DBLP:journals/tkde/Christen12,ElmagarmidIV07} have been mostly conceived for processing structured data of few entity types after being adequately pre-processed in a data warehouse, and hence been able to discover blocking keys of entities and/or mapping rules between their types. We argue that ER techniques are challenged when more than one of the Big Data “V”s have to be addressed simultaneously (e.g., \textit{Volume} or \textit{Velocity} with \textit{Variety}). 

In essence, the high \textit{Variety} of Big Data entities calls for a paradigm shift in all major tasks of ER. Regarding \emph{Blocking}, Variety renders inapplicable the traditional techniques that rely on schema and domain knowledge to maximize the number of comparisons that can be skipped, because they do not lead to matches \cite{DBLP:journals/pvldb/0001APK15}. As far as \emph{Matching} is concerned, Variety requires novel entity matching approaches that go beyond approximate string similarity functions~\cite{Koudas:2006:RLS:1142473.1142599}. This is because such functions are applied on the values of specific attributes among pairs of descriptions, which are difficult to be known in advance. Clearly, \emph{schema-aware} comparisons cannot be used for \emph{loosely structured and highly heterogeneous entity descriptions}, such as those found in LOD. Similarity evidence of entities can be obtained only by looking at the bag of literals contained in descriptions, regardless of the attributes they appear as values. Finally, as the \textit{value-based} similarity of a pair of entities may still be weak due to \textit{Veracity}, we need to consider additional sources of matching evidence related to the \emph{similarity of neighboring} entities, which are connected via relations. 

The previous challenges are exemplified in Figure~\ref{fig:LOD++example}(b), which depicts the two types of similarity for entities known to match from four established benchmark datasets: Restaurant\footnote{ \url{http://oaei.ontologymatching.org/2010/im}}, Rexa-DBLP\footnote{\url{http://oaei.ontologymatching.org/2009/instances}}, BBCmusic-DBpedia\footnote{\url{http://datahub.io/dataset/bbc-music}, \url{http://km.aifb.kit.edu/projects/btc-2012}} and YAGO-IMDb\footnote{\url{http://www.yago-knowledge.org}, \url{http://www.imdb.com}}. Every dot corresponds to a different matching pair, while its shape denotes the respective dataset. The horizontal axis reports the normalized value similarity based on the common words in a pair of descriptions (weighted Jaccard~\cite{DBLP:conf/kdd/Lacoste-JulienPDKGG13}), while the vertical one reports the maximum value similarity of their respective entity neighbors. We can observe that the value-based similarity of matching entities significantly varies across different datasets. For \textit{strongly similar entities} (e.g., 
value similarity $>$ 0.5), existing duplicate detection techniques work well, but to resolve \textit{nearly similar entities} (e.g., value similarity $<$ 0.5), we need advanced ways of exploiting evidence about the similarity of neighboring entities, due to the Variety in entity types.

Additional challenges are raised by the \textit{Velocity} of Big Data Entities. Even though ER is historically framed as an offline task that improves data quality in data warehouses upon completion of data integration, many services now require to \emph{resolve entities in real-time}. Such services strive for incremental ER workflows over \emph{dynamic sources} that can sacrifice completeness of the resulting matches as long as \emph{query-based}~\cite{DBLP:journals/jair/BhattacharyaG07,DBLP:journals/pvldb/AltwaijryMK15} or \emph{streaming}~\cite{DBLP:conf/edbt/KarapiperisGV18} execution strategies can be supported. 

\vspace{3pt}
\noindent
\textbf{Contributions.} 
Record linkage and deduplication techniques for structured data in data warehouse settings are the subject of numerous surveys and benchmarking efforts \cite{DBLP:books/daglib/0030287,DBLP:journals/tkde/Christen12,ElmagarmidIV07,DBLP:journals/pvldb/HassanzadehCML09,DBLP:journals/pvldb/KopckeTR10,DBLP:series/synthesis/2010Naumann,DBLP:books/acm/IlyasC19,DBLP:conf/semweb/EfthymiouHRC17}.  Approximate instance matching is surveyed in~\cite{Dorneles2011}, 
link discovering algorithms in~\cite{DBLP:journals/semweb/NentwigHNR17}, and uncertain ER
in~\cite{DBLP:journals/pvldb/Gal14}. Recent efforts to enhance scalability of ER methods by leveraging distribution and parallelization techniques are surveyed in~\cite{DBLP:journals/ojbd/ChenSS18}, while overviews of blocking and filtering techniques are presented in \cite{o2019review,DBLP:journals/corr/abs-1905-06167}. In contrast, our goal is to present an in-depth survey on all tasks required to implement complex ER workflows, including Indexing, Matching and Clustering.

To the best of our knowledge, this is the first survey that provides an end-to-end view of ER workflows for Big Data entities and of the new entity methods addressing the \textit{Variety} in conjunction with the \textit{Volume} or the \textit{Velocity} of Big Data Entities. Throughout this survey, we present the basic concepts, processing tasks and execution strategies required to cope with the loose \emph{structuredness}, extreme structural \emph{diversity}, high \emph{speed} and large \emph{scale} of entity descriptions actually consumed by Big Data applications. 
This survey is intended to provide a starting point for researchers, students and developers interested in recent advances of schema-agnostic, budget-aware and incremental ER techniques that resolve nearly similar entity descriptions published by numerous Big Data sources. 

The remaining of this survey is organized as follows. Section \ref{erworkflows} presents the core concepts and tasks for building end-to-end ER workflows. Each workflow task is then examined in a separate section: Blocking in Section \ref{sec:blocking}, Block Processing in Section \ref{sec:blockProcessing}, Matching in Section \ref{sec:matching}, and Clustering in Section \ref{sec:clustering}. All these sections study methods for batch ER, while budget-aware and incremental ER are described in Sections \ref{sec:progressiveER} and \ref{sec:incrementalER}, respectively. Section \ref{sec:other} covers complementary ER methods along with the main systems for end-to-end ER, while Section \ref{sec:directions} elaborates on the most important directions for future work. Finally, Section \ref{sec:conclusions} summarizes the current status of ER research.

Note that two of the authors have also published a survey on blocking and filtering (similarity join) techniques for structured and semi-structured data \cite{DBLP:journals/corr/abs-1905-06167}, which covers only two steps of the end-to-end ER workflow for Big Data entities - Blocking in Section \ref{sec:blocking} and Block Processing in Section \ref{sec:blockProcessing}. In contrast, this survey covers the entire end-to-end ER workflow, including Entity Matching, Clustering, and topics such as budget-aware, incremental, crowd-sourced, rule-based, deep learning-based and temporal ER. The overlap of the two surveys is kept to the minimum.

\section{ER Processing Tasks and Workflows}
\label{erworkflows} 

The core notion of \textit{entity description} comprises a set of attribute-value pairs uniquely identified through a global id. A set of such descriptions is called \textit{entity collection}. Two descriptions that are found to correspond to the same real-word object are called \textit{matches} or \textit{duplicates}. Depending on the input and its characteristics, the ER problem is distinguished into \cite{DBLP:conf/edbt/Efthymiou0SC19,DBLP:journals/tkde/PapadakisIPNN13,DBLP:journals/csimq/SaeediNPR18,DBLP:journals/pvldb/SimoniniBJ16}:
\begin{enumerate}
    \item \textit{Clean-Clean ER}, when the input comprises two overlapping, but individually clean (i.e., duplicate-free) entity collections and the goal is to find the matches between them.
    \item \textit{Dirty ER}, where the goal is to identify the duplicates within a single entity collection.
    \item \textit{Multi-source ER}, when more than two entity collections are given as input. 
\end{enumerate}

\begin{figure*}\centering
\includegraphics[width=0.99\textwidth]{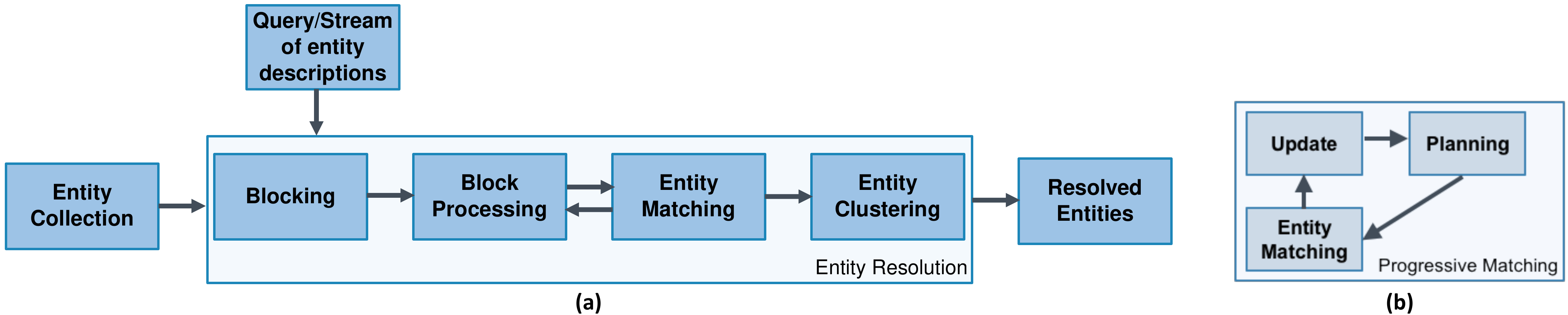}
\vspace{-8pt}
\caption{(a) The generic end-to-end workflow for Entity Resolution. (b) Budget-aware Matching.}
\label{fig:workflow}
\vspace{-10pt}
\end{figure*}

All previous instances of the ER problem involve general processing tasks as illustrated in the end-to-end workflow of Figure~\ref{fig:workflow}(a)~\cite{DBLP:series/synthesis/2015Christophides,DBLP:conf/www/StefanidisEHC14}. 
As every description should be compared to all others, the ER problem is by nature quadratic to the size of the input entity collection(s). To cope with large Volumes of entities, \emph{Blocking} (a.k.a., \emph{Indexing}) is typically applied as a first processing task to discard as many comparisons as possible without missing any matches. It places similar descriptions into blocks, based on some criteria (typically, called \emph{blocking keys}) so that it suffices to execute comparisons only between descriptions co-occurring in at least one block.
In other words, Blocking discards comparisons between descriptions that are unlikely to match, quickly splitting the input entity collection into blocks as close as possible to the final ER result. 

To address Variety in Big Data, Blocking operates in a schema-agnostic fashion that considers all attribute values, regardless of the associated attribute names \cite{DBLP:journals/pvldb/0001SGP16}. The key is \emph{redundancy}, i.e., the act of placing every entity into multiple blocks, thus increasing the likelihood that matching entities co-occur in at least one block. On the flip side, the number of executed comparisons is extremely big. This is addressed, though, by a second processing task, called \textit{Block Processing}. Its goal is to restructure an existing block collection so as to minimize the number of comparisons, without any significant impact on the duplicates that co-occur in blocks. This is achieved by discarding two types of unnecessary comparisons: the \textit{redundant} ones, which are repeated across multiple blocks and the \textit{superfluous} ones, which involve non-matching entities.

The next task is \emph{Matching}, which, in its simplest form,
applies a function $M$ that maps each pair of entity descriptions $(e_i, e_j)$ to $\{true, false\}$, with $M(e_i,e_j) = true$ meaning that $e_i$ and $e_j$ are matches, and $M(e_i, e_j) = false$ that they are not. Typically, the match function is defined via a similarity function $sim$ that measures how similar two descriptions are to each other, according to certain comparison criteria. Finding a similarity function that perfectly distinguishes all matches from non-matches for all entity collections is rather hard. Thus, in reality, we seek a similarity function that is only good enough, minimizing the number of false positive or negative matches. 

Recent works have also proposed an \textit{iterative ER process}, which interleaves Matching with Blocking \cite{DBLP:journals/pvldb/RastogiDG11,WhangMKTG09}: Matching is applied to the results of (Meta-)Blocking and the results of each iteration potentially alter the existing blocks, triggering a new iteration. The block modifications are based on the relationships between the matched descriptions and/or on the results~of~their~merging.  

The final task in the end-to-end ER workflow is \emph{Clustering} \cite{DBLP:journals/pvldb/HassanzadehCML09,DBLP:conf/icdm/NentwigGR16,DBLP:journals/csimq/SaeediNPR18,DBLP:conf/adbis/SaeediPR17,DBLP:conf/esws/SaeediPR18}, which groups together the identified matches such that all descriptions within a cluster match. Its goal is actually to infer indirect matching relations among the detected pairs of matching descriptions so as to overcome possible limitations of the employed similarity functions. Its output comprises disjoint sets of entity descriptions $R = \{r_1, r_2, \ldots, r_m\}$ , such that: $(i)$ $\forall e_i,$ $e_j \in r_k \; M(e_i, e_j)$ = $true$, $(ii)$ $\forall e_i \in r_k \forall e_j \in r_l \; M(e_i, e_j) = false$, and $(iii)$ $\cup_{r_i} r_i \in R = \mathcal{E}$, where $\mathcal{E}$ stands for the input entity collection. This partitioning corresponds to the resulting set of resolved entities in Figure~\ref{fig:workflow}(a).

Figure \ref{fig:workflow}(b) illustrates the additional processing tasks that are required when an ER workflow is subject to budget restrictions in terms of time or number of comparisons. These restrictions essentially call for an approximate solution to ER, as an indirect way of addressing Volume. 
Rather than finding all entity matches, the goal of \emph{budget-aware ER} is to progressively identify as many matches as possible within a specified cost budget. It extends batch, budget-agnostic ER workflows with a \emph{Planning} and \emph{Update} phase that typically work on  windows~\cite{DBLP:journals/pvldb/AltowimKM14}. Planning is responsible for selecting \textit{which} pairs of descriptions will be compared for matching and in \textit{what order}, based on the cost/benefit trade-off. Within every window, it essentially favors the more promising comparisons, which are more likely to increase the targeted benefit (e.g., the number of matches) in the remaining budget. Those comparisons are performed first in the current window and thus, a higher benefit is achieved as early as possible. The Update phase takes into account the results of Matching, such that Planning in a subsequent window will promote the comparison of pairs influenced by the previous matches. This iterative ER process continues until the budget is exhausted. Both phases rely on a graph of dependencies among descriptions~\cite{DBLP:conf/sigmod/DongHMN05}, which leverages budget-agnostic~blocking~methods.

Finally, to resolve in real time entities provided as queries against a known entity collection, or arriving in high Velocity streams, \emph{incremental ER} workflows should be supported. In the first case, a \emph{summarization} of the entity collection can reduce the number of comparisons between a query description and an indexed entity collection, by keeping - ideally in memory - representative entity descriptions for each set of already resolved descriptions~\cite{DBLP:conf/edbt/KarapiperisGV18}. Thus, each query (description) corresponds either to descriptions already resolved to a distinct real-world entity, or to a new one, if it does not match with any other description \cite{DBLP:journals/jair/BhattacharyaG07,SismanisWFHR09,Welch:2012:FAI:2396761.2398719}. To boost time efficiency, ER workflows should support \emph{dynamic indexing/blocking} at varying latencies and thus be able to compare only a small number of highly similar candidate pairs arriving in a streaming fashion. Fast algorithms are also required to incrementally cluster the graph formed by the matched entities in a way that approximates the optimal performance of correlation clustering~\cite{Gruenheid:2014:IRL:2732939.2732943}.

\begin{figure*}[t]
	\center \includegraphics[width=0.7\linewidth]{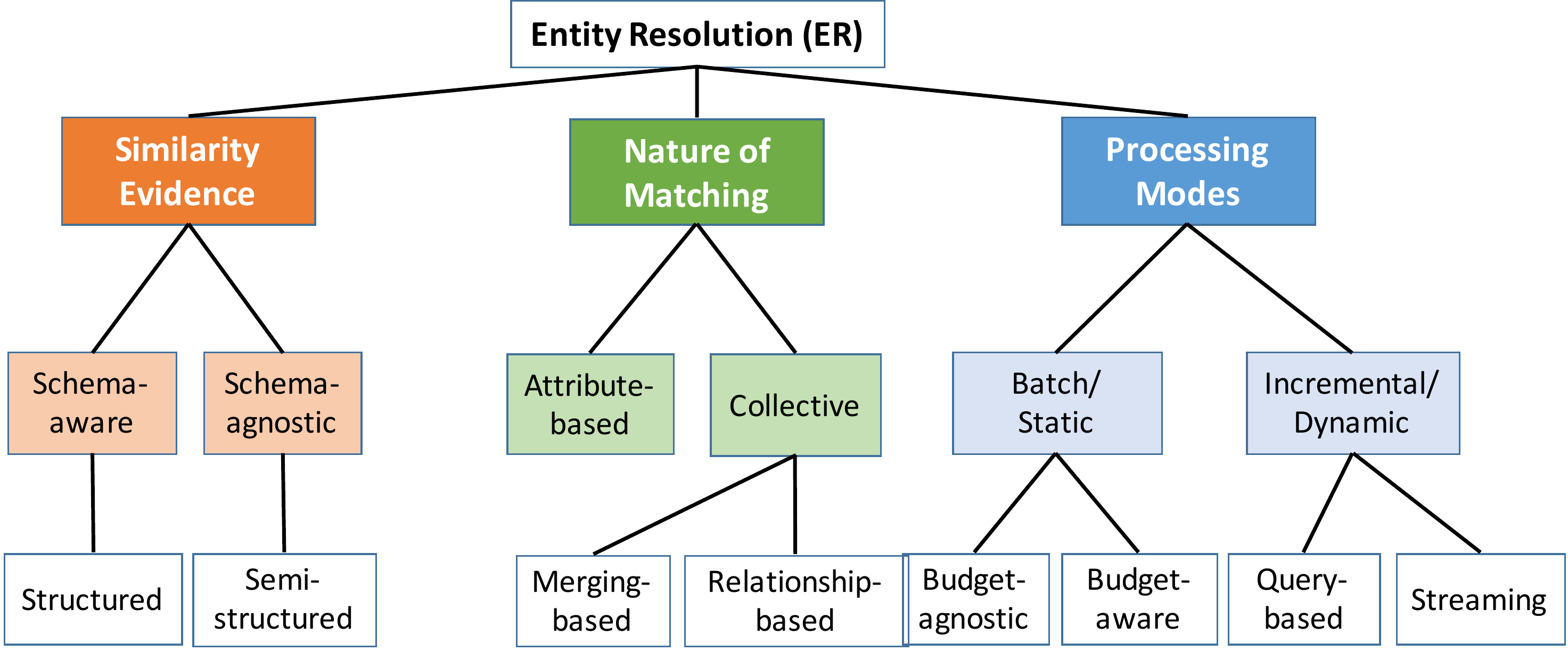}
	\vspace{-10pt}
	\caption{Taxonomy of ER settings and approaches.}
	\label{fig:ERRoadmap}
	\vspace{-10pt}
\end{figure*}

\vspace{3pt}
\noindent
\textbf{Taxonomy of ER settings and approaches.} 
Overall, Figure~\ref{fig:ERRoadmap} illustrates the taxonomy of ER settings based on the key characteristics. Blocking, Matching and Clustering methods that operate on relational data are \emph{schema-aware}, as opposed to the \emph{schema-agnostic} methods, which are more flexible regarding the structure, since they consider all attribute values. In the context of Big Data, nearly similar entities are resolved by going beyond \emph{attribute-based} ER techniques, which examine each pair of descriptions independently from other pairs. To match graph-based descriptions of real-world entities, \emph{collective} ER techniques~\cite{DBLP:journals/tkdd/BhattacharyaG07} are used to increase their matching evidence either by merging partially matched descriptions of entities or by propagating their similarity to neighbor entities via relations that will be matched in a next round. These techniques involve several iterations until they converge to a stable ER result (i.e., no more matches are identified). Thus, collective ER is hard to scale, especially in a cross-domain setting that entails a very large number of sources and entity types. Finally, we distinguish between \emph{batch} (or \emph{static}) ER, which operates on a given input entity collection, and \emph{incremental} (or \emph{dynamic}) ER, which operates on entities arriving in streams or provided by users online as queries. A fine-grained classification of the previous ER settings and approaches will be presented in the following subsections.

\section{Blocking}
\label{sec:blocking}

This step receives as input one or more entity collections and returns as output a set of blocks $\mathcal{B}$, called \textit{block collection}, which groups together similar descriptions, while keeping apart the dissimilar ones. As a result, each description can be compared only to others placed within the same block(s), thus reducing the computational cost of ER to the comparison of similar descriptions. Blocking is thus crucial for successfully addressing the Volume of Big Data.

The desiderata of Blocking are \cite{DBLP:journals/tkde/Christen12}: (i) to place all matching descriptions in at least one common block, and (ii) to minimize the number of suggested comparisons.
The second goal dictates skipping many comparisons, possibly leading to many missed matches, which hampers the first goal. Therefore, Blocking should achieve a good trade-off between these two competing goals.

In this survey, we provide an overview of Blocking for semi-structured data, which require no domain or schema knowledge - unlike the schema-aware methods that are crafted for structured data (we refer the interested reader to \cite{DBLP:journals/corr/abs-1905-06167,DBLP:journals/tkde/Christen12,DBLP:books/daglib/0030287} for more details).
Instead of relying on human intervention, they require no expertise to identify the best attribute(s) for defining blocking keys. They operate in a \textit{schema-agnostic} way that disregards the semantic equivalence of attributes, thus being inherently crafted for addressing the Variety of highly heterogeneous semi-structured data. We distinguish them into non-learning and learning-based methods.

\vspace{3pt}
\noindent
\textbf{Non-learning methods.} \textit{Semantic Graph Blocking} \cite{DBLP:conf/ideas/NinMML07} considers exclusively the relations between descriptions, i.e., foreign keys in databases and links in RDF data. For every description $e_i$, it creates a block $b_i$ that contains all descriptions connected with $e_i$ through a path of restricted length, provided that the block size does not exceed a predetermined limit. 

The textual content of attributes is considered by \textit{Token Blocking} (\textbf{\textsf{TB}}) \cite{DBLP:journals/tkde/PapadakisIPNN13}, which creates a block $b_t$ for every distinct attribute value token $t$, regardless of the associated attribute names: two descriptions co-occur in $b_t \in \mathcal{B}$, if they share token $t$ in any of their attribute values. This crude operation yields high recall, due to \textit{redundancy} (i.e., every entity participates in multiple blocks), at the cost of low precision. This is due to the large portion of \textit{redundant comparisons}, which~are~repeated in different blocks, and \textit{superfluous} ones, which involve non-matching entities~\cite{DBLP:journals/pvldb/0001PK14,DBLP:journals/tkde/PapadakisIPNN13,DBLP:journals/pvldb/0001APK15}.

Discarding these two types of comparisons, especially the superfluous ones, we can raise \textsf{TB}'s precision without any (significant) impact on recall. \textit{Attribute Clustering Blocking}~\cite{DBLP:journals/tkde/PapadakisIPNN13} clusters together attributes with similar values and applies \textsf{TB} independently to the values of every attribute cluster.
\textit{RDFKeyLearner}~\cite{DBLP:conf/semweb/SongH11} applies \textsf{TB} independently to the values of automatically selected attributes, which combine high value discriminability with high description coverage.
\textit{TYPiMatch} \cite{DBLP:conf/wsdm/MaT13} clusters the input descriptions into a set of overlapping types and then applies \textsf{TB} independently to the members of each type. Unlike \textsf{TB}, which tokenizes URIs on all their special characters, \textit{Prefix-Infix(-Suffix) Blocking}~\cite{DBLP:conf/wsdm/PapadakisINPN12} uses as blocking keys only the infixes of URIs - the \textit{prefix} describes the domain of the URI, the \textit{infix} is a local identifier, and the optional \textit{suffix} contains details about the format, or a named anchor. 
For example, in \textit{"https://dl.acm.org/journal/csur/authors"}, the prefix is \textit{"https://dl.acm.org/journal"}, the infix is \textit{"csur"} and the suffix is \textit{"authors"}.

Another family of Blocking methods stems from generalizing \textsf{TB}'s functionality to the main schema-aware non-learning techniques. 
By using the same blocking keys as \textsf{TB}, 
we can apply traditional Blocking methods to heterogeneous semi-structured data~\cite{DBLP:journals/pvldb/0001APK15} and significantly improve their recall, even over structured data. This has been successfully applied to the following techniques:

\textit{Suffix Arrays Blocking}~\cite{DBLP:dblp_conf/wiri/AizawaO05} converts each \textsf{TB} blocking key (i.e., attribute value token) into the suffixes that are longer than a specific minimum length $l_{min}$. Then, it defines a block for every suffix that does not exceed a predetermined frequency threshold $b_{max}$, which specifies the maximum block size. \textit{Extended Suffix Arrays Blocking}~\cite{DBLP:journals/tkde/Christen12,DBLP:journals/pvldb/0001APK15} considers all substrings (not just the suffixes) of \textsf{TB} blocking keys with more than $l_{min}$ characters, so as to support noise at the end of blocking keys (e.g., ``JohnSnith" and ``JohnSmith"). Similarly, 
\textit{Q-grams Blocking}~\cite{DBLP:journals/tkde/Christen12,DBLP:journals/pvldb/0001APK15} converts every \textsf{TB} blocking key into sub-sequences of $q$ characters (\textit{$q$-grams}) and defines a block for every distinct $q$-gram. \textit{Extended Q-Grams Blocking} \cite{DBLP:journals/tkde/Christen12,DBLP:journals/pvldb/0001APK15} concatenates multiple $q$-grams to form more distinctive blocking keys.

\textit{Canopy Clustering}~\cite{DBLP:conf/kdd/McCallumNU00,DBLP:journals/tkde/Christen12} iteratively selects a random description $e_i$ and creates a new block $b_i$ for it. Using a cheap string similarity measure, it places in $b_i$ all descriptions whose \textsf{TB} blocking keys have a similarity to $e_i$ higher than $t_{in}$; descriptions with a similarity higher~than~$t_{ex}$($>$ $t_{in}$) participate in no subsequent block. \textit{Extended Canopy Clustering} \cite{DBLP:journals/tkde/Christen12,DBLP:journals/pvldb/0001APK15} replaces the weight thresholds with cardinality ones: for each randomly selected description, the $k_1$ most similar descriptions are placed in its block, while the $k_2 (\leq$ $k_1)$ most similar ones participate in no~other~block. 

Finally, \textit{Sorted Neighborhood} \cite{DBLP:conf/sigmod/HernandezS95} sorts \textsf{TB} blocking keys in alphabetical order. A window of fixed size $w$ slides over the sorted list of descriptions and compares the description at the last position with all descriptions in the same window. This approach is robust to noise in blocking keys, but small $w$ trades high precision for low recall and vice versa for large $w$~\cite{DBLP:journals/tkde/Christen12}. To address this issue, \textit{Extended Sorted Neighborhood} \cite{DBLP:journals/tkde/Christen12,DBLP:journals/pvldb/0001APK15} slides the window $w$ over the sorted list of \emph{blocking keys}. 

\vspace{3pt}
\noindent
\textbf{Learning-based methods.} \textit{Hetero} \cite{DBLP:conf/semweb/KejriwalM14a} is an unsupervised approach that maps every dataset to a normalized TF vector, and applies an efficient adaptation of the Hungarian algorithm 
to produce positive and negative feature vectors. Then, it applies \textit{FisherDisjunctive} \cite{DBLP:conf/icdm/KejriwalM13} with bagging to achieve robust performance. \textit{Extended DNF BSL} \cite{DBLP:journals/corr/KejriwalM15} combines an established instance-based schema matcher with weighted set covering to learn supervised blocking schemes in Disjunctive Normal Form (DNF) with at most $k$ attributes.

\vspace{3pt}
\noindent
\textbf{Parallelization.} Parallel adaptations of the above methods have been proposed in the literature. They rely on the \textit{MapReduce paradigm} \cite{DeanG04}: following a split-apply-combine strategy, MapReduce partitions the input data into smaller chunks, which are then processed in parallel. A \texttt{Map} function emits intermediate (key, value) pairs for each input split, while a \texttt{Reduce} function processes the list of values that correspond to a particular intermediate key, regardless of the mapper that emitted them. The two functions form a MapReduce job, with complex procedures involving multiple jobs.

Using a single MapReduce job, \textsf{TB} builds an inverted index that associates every token with all entities containing it in their attribute values \cite{DBLP:series/synthesis/2015Christophides,DBLP:conf/bigdataconf/EfthymiouSC15}.  For Attribute Clustering, four MapReduce jobs are required \cite{DBLP:series/synthesis/2015Christophides,DBLP:conf/bigdataconf/EfthymiouSC15}: the first one aggregates all values per attribute,
the second one estimates the similarity between all attributes,
the third one associates every attribute with its most similar one, and the fourth one assigns to every attribute a cluster id and applies the \textsf{TB} MapReduce job. Prefix-Infix(-Suffix) Blocking requires three jobs \cite{DBLP:series/synthesis/2015Christophides,DBLP:conf/bigdataconf/EfthymiouSC15}: the first two
extract the prefixes and the optional suffixes from the input URIs, respectively, while the third one applies \textsf{TB}'s mapper to the literal values and a specialized mapper that extracts infixes to the URIs. 

A crucial aspect of the MapReduce paradigm is the \textit{load balancing algorithm}. To balance the cost of executing the comparisons defined in an existing block collection, \textit{Dis-Dedup}~\cite{DBLP:journals/pvldb/ChuIK16} formalizes load balancing as an optimization problem that minimizes not only the computational, but also the communication cost (e.g., network transfer time, local disk I/O time). The proposed solution provides strong theoretical guarantees for a performance close to the optimal one.

\subsection{Discussion}

\begin{table*}[t]\centering
\vspace{-5pt}
\caption{A taxonomy of the Blocking methods discussed in Section \ref{sec:blocking} (in the order of presentation).}
\vspace{-5pt}
\includegraphics[width=0.79\linewidth]{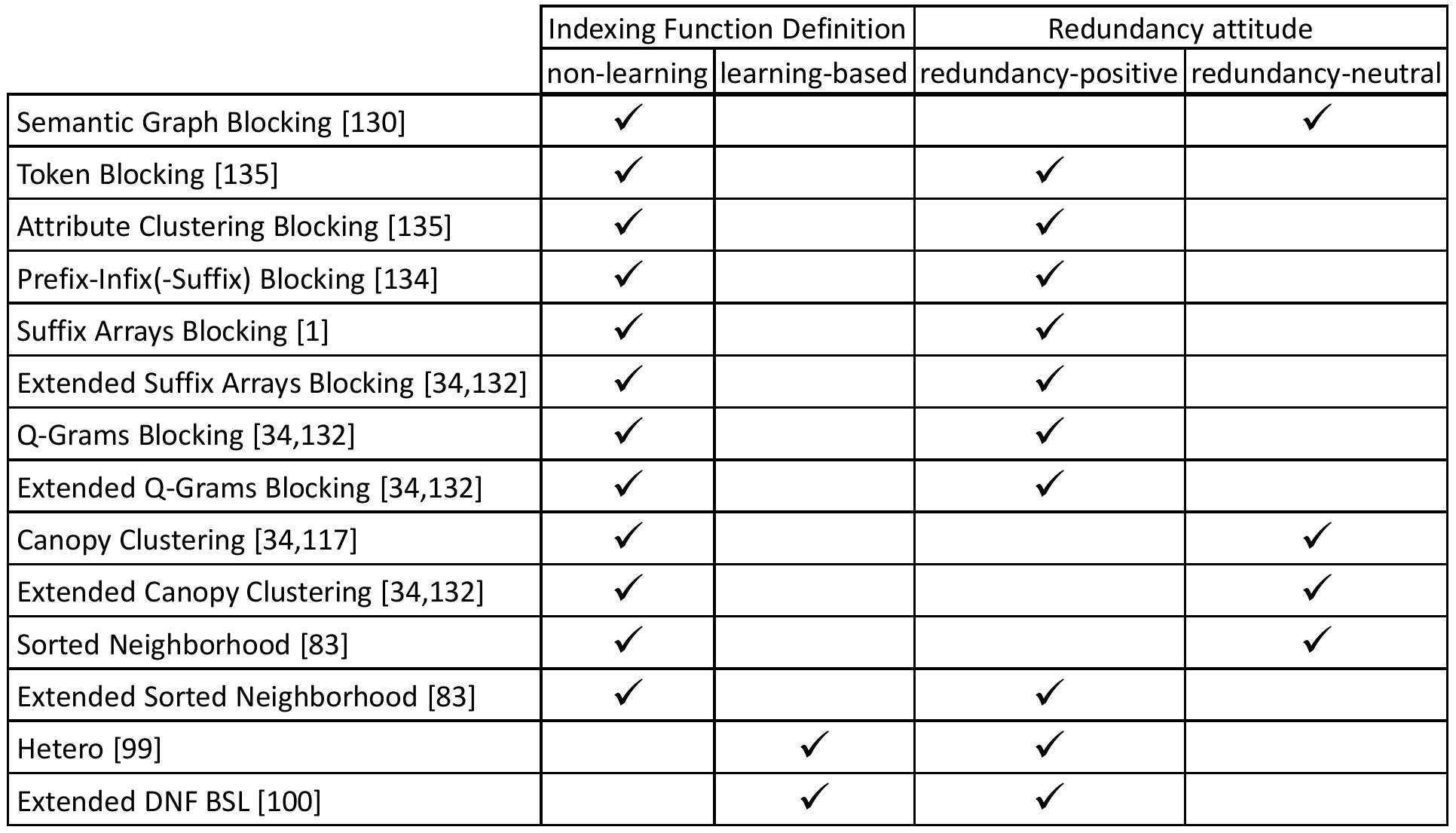}
\label{fig:blockingTaxonomy}
\vspace{-5pt}
\end{table*}

Table \ref{fig:blockingTaxonomy} organizes the main schema-agnostic Blocking methods in a two-dimensional taxonomy that is formed by two criteria: (i) \textit{Indexing Function Definition}, which determines whether learning is used to extract blocking keys from each entity description, and \textit{Redundancy attitude}, which determines whether the outcome is a \textit{redundancy-positive block collection}, where the more blocks two descriptions share, the more likely they are to be matching, or a \textit{redundancy-neutral one} otherwise. We observe that most methods involve a non-learning functionality that produces redundancy-positive blocks. Among them, \textsf{TB} tries to maximize recall by assuming 
that duplicate entities share at least one common token in their values. 
Extensive experiments have shown that this assumption holds for KBs in the \textit{center of the LOD cloud}~\cite{DBLP:series/synthesis/2015Christophides,DBLP:conf/bigdataconf/EfthymiouSC15}. Yet, this coarse-grained approach typically leads to very low precision, since most of the pairs sharing a common word are non-matches. 
Attribute Clustering Blocking increases \textsf{TB}'s precision by requiring that the common tokens of matching entities appear in attributes with similar values. 
Prefix-Infix(-Suffix) Blocking applies only to RDF data. However, it has been shown that both methods perform poorly when applied to KBs from the \textit{periphery of the LOD cloud}~\cite{DBLP:conf/bigdataconf/EfthymiouSC15,DBLP:series/synthesis/2015Christophides}. The reason is that they exclusively consider the noisy content of descriptions, disregarding the valuable evidence that is provided by contextual information, such as the neighboring descriptions, i.e., entities of different types connected via important relations. TYPiMatch also attempts to raise \textsf{TB}'s precision, by categorizing the given entities into overlapping types, but its recall typically drops to a large extent, due to the noisy, schema-agnostic detection of entity types \cite{DBLP:journals/pvldb/0001SGP16}. 

Overall, the schema-agnostic Blocking methods address both Volume and Variety of Big Data entities, consistently achieving high recall, due to redundancy.
Their precision, though, is very low, due to the large portion of redundant and the superfluous comparisons in their overlapping blocks. We refer to \cite{DBLP:journals/corr/abs-1905-06167,DBLP:journals/tkde/Christen12,DBLP:books/daglib/0030287} for a more detailed overview of Blocking methods.

\vspace{-5pt}

\section{Block Processing}
\label{sec:blockProcessing}

This step receives as input a set of blocks $\mathcal{B}$ and produces as output a new set of blocks $\mathcal{B'}$ that has similar recall, but significantly higher precision. This is achieved by discarding most superfluous and redundant comparisons in $\mathcal{B}$. The relevant techniques operate at the coarse level of entire blocks (Block Cleaning) or at the finer level of individual comparisons (Comparison Cleaning). 

\subsection{Block Cleaning}
\label{sec:blcl}

Methods of this type are \textit{static}, i.e., independent of Matching, or \textit{dynamic}, i.e., interwoven with it. 

\vspace{3pt}
\noindent
\textbf{Static methods.} The core assumption is that excessively large blocks (e.g., those corresponding to stop-words) are dominated by unnecessary comparisons. In fact, the larger a block is, the less likely it is to contain \textit{unique duplicates}, i.e., matches that share no other block. Hence, they discard the largest blocks, raising precision, without any significant impact on recall. To this end, 
\textit{Block Purging} sets an upper limit on the number of comparisons \cite{DBLP:journals/tkde/PapadakisIPNN13} or the block size \cite{DBLP:conf/wsdm/PapadakisINPN12}. \textit{Block Filtering} applies a limit to the blocks of every description, retaining it in $r\%$ of its~smallest~blocks~\cite{DBLP:conf/edbt/0001PPK16,DBLP:journals/pvldb/0001SGP16}.

More advanced methods, like a MapReduce-based blocking algorithm~\cite{mcneill2012dynamic}, learning-based (supervised) method \textit{Rollup Canopies}~\cite{DBLP:conf/cikm/SarmaJMB12} and \textit{Size-based Block Clustering}~\cite{DBLP:conf/kdd/FisherCWR15}, split excessively large blocks into smaller sub-blocks until they all satisfy the maximum block size limit. 
The last method may merge back small blocks with similar blocking keys, in order to raise recall.

\vspace{3pt}
\noindent
\textbf{Dynamic methods.}
Assuming that Matching is performed by a \textit{perfect oracle}, these methods schedule the processing of blocks on-the-fly so as to maximize ER effectiveness and time efficiency.
For Dirty ER, \textit{Iterative Blocking} \cite{WhangMKTG09} merges any new pair of matching descriptions, $e_i$ and $e_j$, into a new one, $e_{i,j}$ and replaces both $e_i$ and $e_j$ with $e_{i,j}$ in all blocks that contain them. The already processed blocks are reprocessed so that $e_{i,j}$ is compared with all others; the new content in $e_{i,j}$ may yield different similarity values that designate previously missed matches. 

For Clean-Clean ER, \textit{Block Scheduling} orders blocks in ascending order of comparisons~\cite{simonini2018schema}, or block size~\cite{DBLP:journals/tkde/PapadakisIPNN13}, so as to detect matches as early as possible. These matches are propagated to subsequently processed blocks in order to reduce the superfluous comparisons. This yields a block processing order with decreasing density of detected matches. Based on this observation, \textit{Block Pruning}~\cite{DBLP:journals/tkde/PapadakisIPNN13} terminates the entire ER process as soon as the average number of executed comparisons for detecting a new pair of duplicates drops below a predetermined threshold.

\subsection{Comparison Cleaning}
\label{sec:cocl}

Most methods of this type operate on \textit{redundancy-positive block collections}, where the more blocks two descriptions share, the more likely they are to be matching. This characteristic allows for weighting all pairwise comparisons in proportion to the matching likelihood of the corresponding descriptions, a process that has been formalized by \textit{Meta-blocking}~\cite{DBLP:journals/tkde/PapadakisKPN14}.

Meta-blocking converts the input block collection $\mathcal{B}$ into a \textit{blocking graph} $G_B$, where nodes correspond to descriptions and unique edges connect every pair of co-occurring descriptions. The edges are weighted in proportion to the likelihood that the adjacent descriptions are matching. 
Edges with low weights are pruned, as they probably correspond to superfluous comparisons. A new block is then created for every retained edge, yielding the restructured block collection $\mathcal{B}'$. In this process, various techniques can be used for weighting and pruning the graph edges~\cite{DBLP:journals/tkde/PapadakisKPN14}. 

For edge pruning, the following algorithms are available: \textit{Weighted Edge Pruning}~\cite{DBLP:journals/tkde/PapadakisKPN14} removes all edges that do not exceed the average edge weight; \textit{Cardinality Edge Pruning} retains the globally $K$ top weighted edges \cite{DBLP:journals/tkde/PapadakisKPN14,zhang2017pruning}; \textit{Weighted Node Pruning} (\textsf{WNP})~\cite{DBLP:journals/tkde/PapadakisKPN14} and \textit{BLAST}~\cite{DBLP:journals/pvldb/SimoniniBJ16} retain in each node neighborhood the descriptions that exceed a local threshold; \textit{Cardinality Node Pruning} (\textsf{CNP}) retains the top-$k$ weighted edges in each node neighborhood~\cite{DBLP:journals/tkde/PapadakisKPN14}; \textit{Reciprocal WNP} and \textit{CNP}~\cite{DBLP:conf/edbt/0001PPK16} retain edges satisfying the pruning criteria in both adjacent node neighborhoods. Other methods perform edge pruning inside individual blocks~\cite{DBLP:journals/kais/NascimentoPM20}, while \textit{Disjunctive Blocking Graph} \cite{DBLP:conf/edbt/Efthymiou0SC19} associates every edge with multiple weights to express composite co-occurrence conditions.

On another line of research, \textit{Transitive LSH}~\cite{DBLP:conf/psd/SteortsVSF14} converts \textsf{LSH} blocks into an unweighted blocking graph and applies a community detection algorithm, such as~\cite{clauset2004finding}, while \textit{SPAN}~\cite{DBLP:conf/icde/ShuCXM11} uses matrix representations and operations to enhance the input block collection. The only approach that applies to any block collection $\mathcal{B}$, even one that is not redundancy-positive, is \textit{Comparison Propagation}~\cite{DBLP:journals/tkde/PapadakisIPNN13}, 
which merely discards all redundant comparisons from~$\mathcal{B}$. 

\begin{table*}[t]\centering
\vspace{-5pt}
\caption{A taxonomy of the Blocking Processing methods discussed in Section \ref{sec:blockProcessing} (in the order of presentation).}
\vspace{-5pt}
\includegraphics[width=0.79\linewidth]{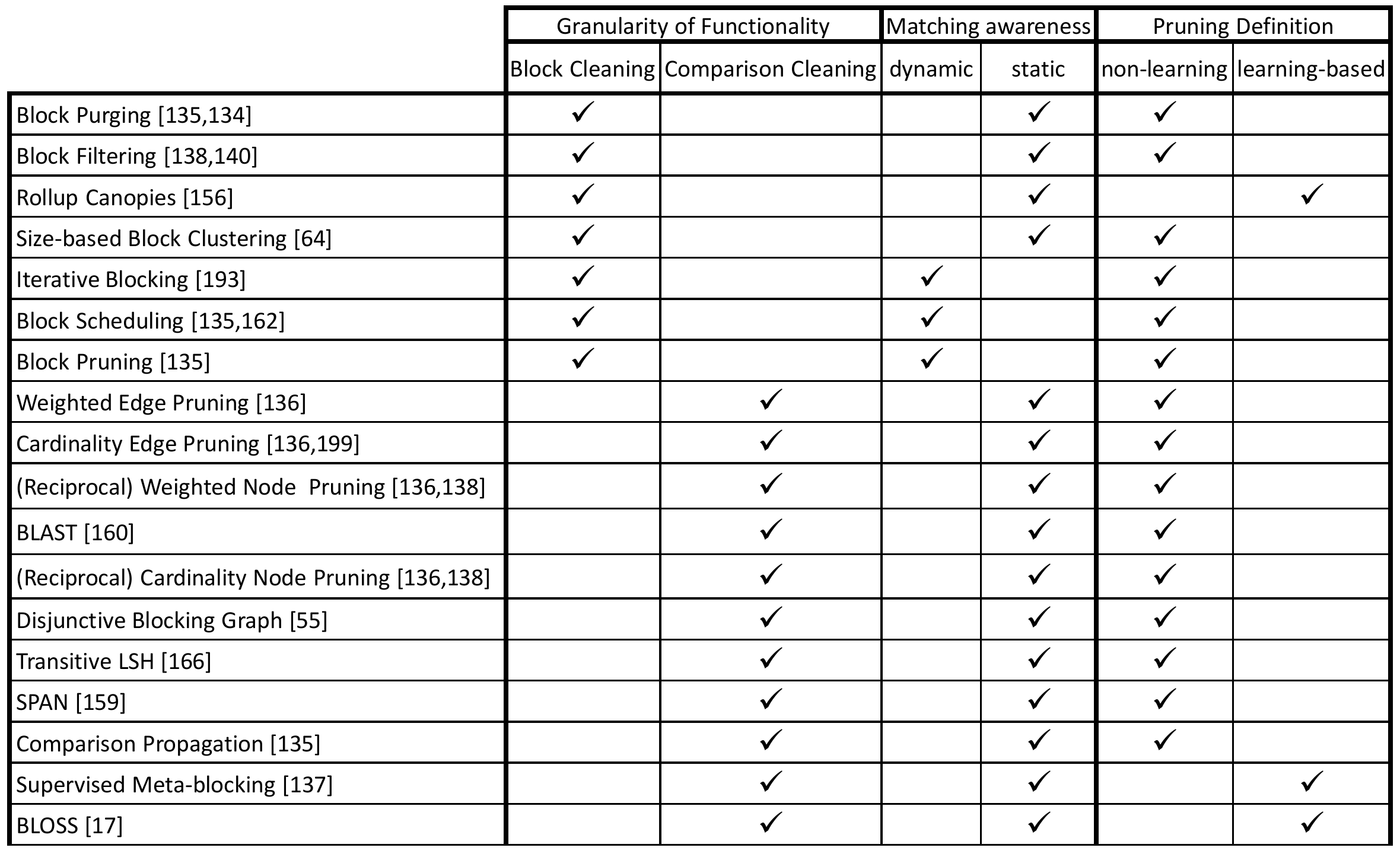}
\label{fig:blockProcessingTaxonomy}
\vspace{-5pt}
\end{table*}

\vspace{3pt}
\noindent
\textbf{Learning-based methods.}
\textit{Supervised Meta-blocking}~\cite{DBLP:journals/pvldb/0001PK14} casts edge pruning as a binary classification problem: every edge is annotated with a vector of schema-agnostic features, and is classified as \texttt{likely match} or \texttt{unlikely match}. \textit{BLOSS}~\cite{DBLP:journals/is/BiancoGD18} further cuts down on the labelling effort, by selecting a very small training set that maintains high effectiveness.

\vspace{3pt}
\noindent
\textbf{Parellelization.} Meta-blocking has been adapted to both multi-core~\citep{DBLP:conf/i-semantics/0001BPK17} and MapReduce parallelization~\citep{DBLP:journals/is/Efthymiou0PSP17}. 
Regarding the latter, the \textit{entity-based strategy}~\citep{DBLP:journals/is/Efthymiou0PSP17} 
aggregates for every description the bag of all description ids that co-occur with it in at least one block. Then, it estimates the edge weight that corresponds to each neighbor based on its frequency in the co-occurrence bag. An alternative approach is the \textit{comparison-based strategy}~\citep{DBLP:journals/is/Efthymiou0PSP17}: the first pre-processing job enriches each block with the list of block ids associated with every description. This allows for computing the edge weights and discarding all redundant comparisons in the Map phase of the second job, while the superfluous comparisons are pruned in the ensuing Reduce phase. 
Both strategies rely on the load balancing algorithm \textit{MaxBlock}~\citep{DBLP:journals/is/Efthymiou0PSP17} to avoid the underutilization of the available resources. BLAST is parallelized in~\citep{DBLP:journals/is/SimoniniGBJ19}, exploiting the broadcast join of Apache Spark for very high efficiency.

\subsection{Discussion}

Table \ref{fig:blockProcessingTaxonomy} presents an overview of the Block Processing methods discussed above. The resulting taxonomy consists of three criteria: granularity of functionality, matching awareness (i.e., whether a method is dynamic, depending on the outcomes of Entity Matching method, or static) and pruning definition (i.e., whether the search space is reduced through a learning process that involves labelled instances or not). Most Block Processing techniques involve a comparison-centric, static and non-learning functionality that can be seamlessly combined with any Blocking technique. 
Numerous studies have demonstrated that Block and Comparison Cleaning are indispensable for schema-agnostic Blocking, raising precision by orders of magnitude, without hurting recall~\cite{DBLP:journals/pvldb/0001SGP16,DBLP:journals/tkde/PapadakisIPNN13,DBLP:journals/pvldb/SimoniniBJ16}.
Multiple Block Cleaning methods can be part of the same end-to-end ER workflow, as they are typically complementary; e.g., Block Purging is usually followed by Block Filtering~\cite{DBLP:conf/edbt/0001PPK16}. Yet, at most one Comparison Cleaning method can be part of an ER workflow: applying it to a redundancy-positive block collection removes its co-occurrence patterns and renders all other techniques inapplicable. The top performer among non-learning techniques is BLAST~\cite{DBLP:journals/pvldb/SimoniniBJ16}, while BLOSS performs better by labelling just $\sim$50 instances~\cite{DBLP:journals/is/BiancoGD18}. We refer to \cite{DBLP:journals/corr/abs-1905-06167} for a more detailed overview of Block Processing techniques.

\section{Matching}
\label{sec:matching}

At the core of ER lies the \emph{Matching} task, which receives as input a block collection and for each pair of candidate matches that co-occur in a block, it  decides if they refer to the same real-world entity.

\subsection{Preliminaries}
\label{ssec:matching_preliminaries}

The matching decision is typically made by a match function $M$, which maps each pair of entity descriptions $(e_i, e_j)$ to $\{true, false\}$, with $M(e_i,e_j) = true$ meaning that $e_i$ and $e_j$ are matches, and $M(e_i, e_j) = false$ meaning that $e_i$ and $e_j$ are not matches. 

In its simplest form, $M$ is defined via a similarity function $sim$, measuring how similar two entities are to each other, according to certain comparison attributes. $sim$ can consist of an \textit{atomic} similarity measure, like Jaccard similarity, or a \textit{composite} one, e.g., a linear combination of several atomic similarity functions on different attributes of a description.  
To specify an equivalence relation among entity descriptions, we need to consider a similarity measure satisfying the non-negativity, identity, symmetry and triangle inequality properties~\cite{Zezula:2010:SSM:1951721}, i.e., a similarity \emph{metric}.  Given a similarity threshold $\theta$, a simple matching function can be defined as:
{\small
\[ 
M(e_i, e_j) = 
	\begin{cases}
	\text{true, if } sim(e_i, e_j) \geq \theta, \\
	\text{false, otherwise.} \\
	\end{cases}
\]
}

In more complex ER pipelines, such as when matching rules are manually provided, or learned from training data, the matching function $M$ can be defined as a complex function involving several matching conditions. For instance, two person descriptions match if their SSN is identical, or if their date of birth, zip code and last names are identical, or if their e-mail addresses are identical. 

Finding a similarity metric which can perfectly distinguish all matches from non-matches using simple pairwise comparisons on the attribute values of two descriptions is practically impossible. In particular, similarity metrics are too restrictive to identify nearly similar matches. Thus, in reality, we seek similarity functions that will be only good enough, i.e., minimize the number of misclassified pairs, and rely on collective ER approaches to propagate the similarity of the entity neighbors of two descriptions to the similarity of those descriptions. In this inherently iterative process, the employed match function is based on a similarity that dynamically changes from iteration to iteration, and its results may include a third state, the \emph{uncertain} one. Specifically, given two similarity thresholds $\theta$ and $\theta'$, with $\theta' < \theta$, the match function at iteration $n$, $M^n$, is given by:
{\small
\[ 
M^n(e_i, e_j) = 
	\begin{cases}
	\text{true, if } sim^{n-1}(e_i, e_j) \geq \theta, \\
	\text{false, if } sim^{n-1}(e_i, e_j) \leq \theta', \\
	\text{uncertain, otherwise.} \\
	\end{cases}
\]
}

Based on the characteristics of the entity collections (e.g., structuredness, domain, size), the nature of comparisons (attribute-based or collective), as well as the availability of known, pre-labeled matching pairs, different methodologies can be followed to identify an appropriate similarity function and thus, a fitting match function. In what follows, we explore alternative methodologies for the matching task and discuss the cases in which those methodologies are more suited. 

\subsection{Collective methods}

To minimize the number of missed matches, commonly corresponding to nearly similar matches, a collective ER process can jointly discover matches of inter-related descriptions. This is an inherently iterative process that entails additional processing cost. We distinguish between \textit{merging-} and \textit{relationship-based} collective ER approaches. In the former, new matches can be identified by exploiting the merging of the previously found matches, while in the latter, iterations rely on the similarity evidence provided by descriptions being structurally related in the original entity graph. 

\begin{figure*}[t]
	\vspace{-5pt}
	\center \includegraphics[width=0.99\linewidth]{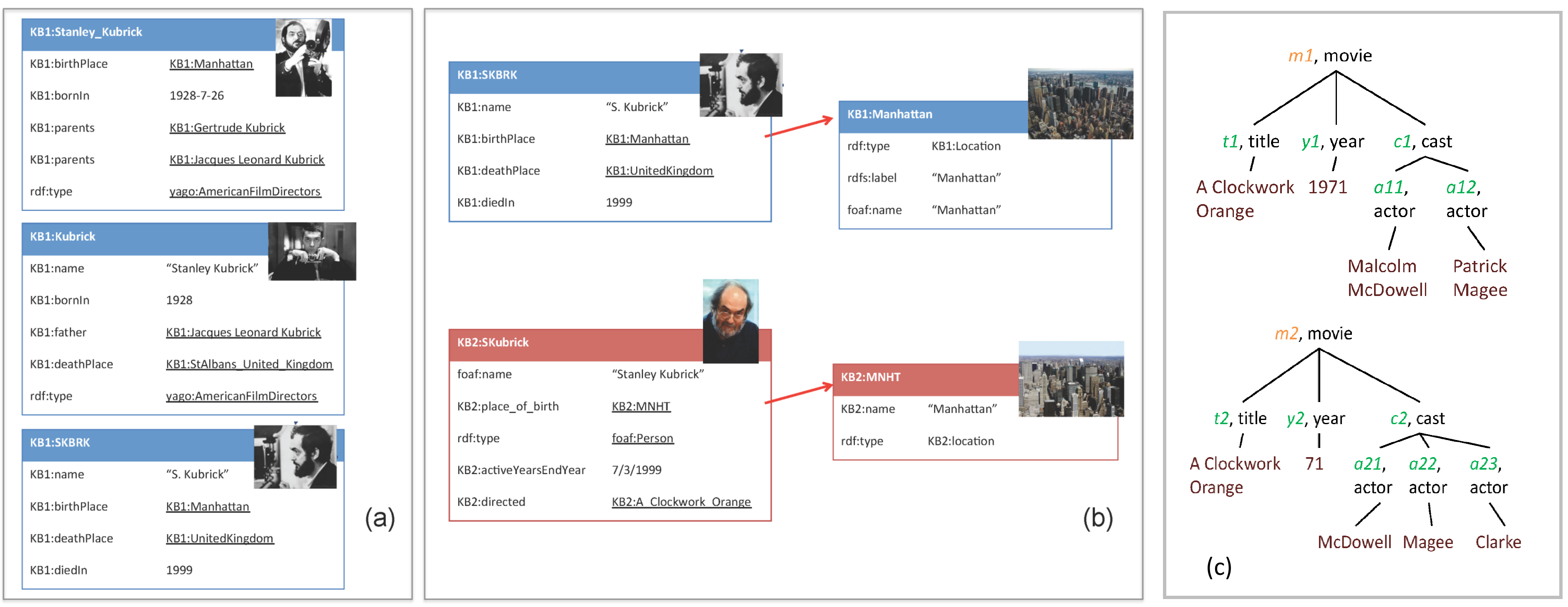}
	\vspace{-12pt}
	\caption{(a) A merging-based collective ER example and (b) a relationship-based collective ER example. 
	(c)  Two different descriptions of the movie \emph{A Clockwork Orange} and its cast in XML. 
	}
	\label{fig:iterativeERexample}
	\vspace{-8pt}
\end{figure*}

\begin{example} Consider the descriptions in Figure~\ref{fig:iterativeERexample}(a), which stem from the knowledge base \emph{KB1}. They all refer to the person Stanley Kubrick. Initially, it is difficult to match \emph{KB1:SKBRK} with any other description, since many people named Kubrick may have been born in Manhattan, or died in the UK, respectively. However, it is quite safe to match the first two descriptions (\emph{KB1:Stanley\_Kubrick} and \emph{KB1:Kubrick}). By merging the first two descriptions, e.g., using the union of their attribute-value pairs, it becomes easier to identify that the last description (\emph{KB1:SKBRK}) also refers to the same person, based on the name and the places of birth and death. Consider now the descriptions in Figure~\ref{fig:iterativeERexample}(b), which stem from the knowledge bases \emph{KB1} and \emph{KB2}. The descriptions on the left (\emph{KB1:SKBRK} and \emph{KB2:SKubrick}) represent Stanley Kubrick, while the descriptions on the right (\emph{KB1:Manhattan} and \emph{KB2:} \emph{MNHT}) represent Manhattan, where Kubrick was born. Initially, it is difficult to identify the match between the descriptions on the left, based only on the common year of death and last name. However, it is quite straightforward to identify the match between the descriptions of Manhattan, on the right. Having identified this match, a relationship-based collective ER algorithm would re-consider matching \emph{KB1:SKBRK} to \emph{KB2:SKubrick}, since these descriptions are additionally related, with the same kind of relationship (birth place), to the descriptions of Manhattan that were previously matched. Therefore, a relationship-based ER algorithm would identify this new match in a second iteration. 
\end{example}

Note that the structuredness of the input entity collection to be resolved is a key factor for the nature of collective approaches. Merging-based methods are typically schema-aware, since structured data make the process of merging easier. On the other hand, collective methods dealing with semi-structured data are typically relationship-based, since merging would require deciding not only which values are correct for a given attribute, but also which values are available for similar attributes and can be used to merge two descriptions. 

\subsubsection{Schema-aware methods} 

In \emph{merging-based collective ER}, the matching decision between two descriptions triggers a merge operation, which transforms the initial entity collection by adding the new, merged description and potentially removing the two initial descriptions. This change also triggers more updates in the matching decisions, since the new, merged description needs to be compared to the other descriptions of the collection. Intuitively, the final result of merging-based collective ER is a new entity collection, which is the result of merging all the matches found in the initial collection. In other words, there should be an 1-1 correspondence between the descriptions in the resolution results and the actual real-world entities from the input entity collection. 

Considering the functions of matching $M$ and merging $\mu$ as black boxes, \textit{Swoosh}~\cite{DBLP:journals/vldb/BenjellounGMSWW09} is a family of merging-based collective ER strategies that minimize the number of invocations to these potentially expensive black boxes; \textit{D-Swoosh}~\cite{DBLP:conf/icdcs/BenjellounGGKLMT07} introduces a family of algorithms that distribute the workload of merging-based ER across multiple processors. 
Both works rely on the following set of $ICAR$ properties, that, when satisfied by $M$ and $\mu$, lead to higher efficiency:

{\small
\noindent$\bullet$\emph{~Idempotence:~} $\forall e_i, M(e_i, e_i) = true$ and $\mu(e_i, e_i)$=$e_i$

\noindent$\bullet$\emph{~Commutativity:~} $\forall e_i, e_j, M(e_i, e_j)$=$true \Leftrightarrow M(e_j, e_i)$= $true$ and $\mu(e_i, e_j) = \mu(e_j, e_i)$

\noindent$\bullet$\emph{~Associativity:~} $\forall e_i, e_j, e_k,$ if both $\mu(e_i, \mu(e_j,e_k))$ and $\mu(\mu(e_i, e_j),e_k)$ exist, $\mu(e_i, \mu(e_j,e_k))$=$\mu(\mu(e_i, e_j),e_k)$

\noindent$\bullet$\emph{~Representativity:~} If $e_k = \mu(e_i, e_j)$, then for any $e_l$ such that $M(e_i,e_l) = true$, $M(e_k, e_l) = true$
}

\noindent
Regarding the match function, idempotence and commutativity have been already discussed in Section~\ref{ssec:matching_preliminaries}, as reflexivity and symmetry, respectively, while representativity extends transitivity, by also including the merge function. Note that if associativity does not hold, it becomes harder to interpret a merged description, since it depends on the order in which the source descriptions~were~merged. 

\textit{R-Swoosh}~\cite{DBLP:journals/vldb/BenjellounGMSWW09} exploits the $ICAR$ properties as follows. A set $\mathcal{E}$ of entity descriptions is initialized to contain all the input descriptions. Then, in each iteration, a description $e$ is removed from $\mathcal{E}$ and compared to each description $e'$ of the, initially empty, set $\mathcal{E'}$. If $e$ and $e'$ are found to match, then they are removed from $\mathcal{E}$ and $\mathcal{E'}$, respectively, and the result of their merging is placed into $\mathcal{E}$ (exploiting representativity). If there is no description $e'$ matching with $e$, then $e$ is placed in $\mathcal{E'}$. This process continues until $\mathcal{E}$ becomes empty, i.e., there are no more matches to be found. 

In \emph{relationship-based collective ER}, the matching decision between two descriptions triggers discovering new candidate pairs for resolution, or re-considering pairs already compared; matched descriptions may be related to other descriptions, which are now more likely to match to each~other.

To illustrate the relationships between the descriptions of an entity collection $\mathcal{E}$, usually, an \textit{entity graph} $G_\mathcal{E}=(V,E)$ is used, in which nodes, $V \subseteq \mathcal{E}$, represent entity descriptions and edges, $E$, reflect the relationships between the nodes. For example, such a match function could be of~the~form:
{\small
$$
M(e_i, e_j) = \left\{ \begin{array}{ll}
						true, \text{ if } sim(nbr(e_i), nbr(e_j)) \geq \theta \\
						false, \text{ else,} \\
						\end{array}
						\right.
$$
}
where $sim$ can be a relational similarity function and $\theta$ is a threshold value. Intuitively, the neighborhood $nbr(e)$ of a node $e$ can be the set of all the nodes connected to $e$, i.e., $nbr(e) = \{e_j | (e, e_j) \in E\}$, or the set of edges containing $e$, i.e., $nbr(e) = \{(e,e_j) | (e,e_j) \in E\}$. 

\textit{Collective ER}~\cite{DBLP:journals/tkdd/BhattacharyaG07} employs an entity graph, following the intuition that two nodes are more likely to match, if their edges connect to nodes corresponding to the same entity. To capture this iterative intuition, hierarchical agglomerative clustering is performed, where, at each iteration, the two most similar clusters are merged, until the similarity of the most similar clusters is below a threshold. When two clusters are merged, the similarities of their related clusters, i.e., the clusters corresponding to descriptions related to the descriptions in the merged cluster, are updated. To avoid comparing all the pairs of input descriptions, Canopy Clustering~\cite{DBLP:conf/kdd/McCallumNU00} is initially applied.

\textit{Hybrid Collective ER}~\cite{DBLP:conf/sigmod/DongHMN05} is based on both partial merging results and relations between descriptions. It constructs a dependency graph, where every node represents the similarity between a pair of entity descriptions and every edge represents the dependency between the matching decisions of two nodes. If the similarity of a pair of descriptions changes, the neighbors of this pair might need a similarity re-computation. The dependencies between the matching decisions are distinguished between Boolean and real-valued. The former suggest that the similarity of a node depends only on whether the descriptions of its neighbor node match or not, while in real-valued dependencies, the similarity of a node depends on the similarity of the descriptions of its neighbor node. Boolean dependencies are further divided into strong (if a node corresponds to a match, its neighbor pair should also be a match), and weak (if a node corresponds to a match, the similarity of its neighbor pair is increased). Initially, all nodes are added to a priority queue. On each iteration, a node is removed from the queue and if the similarity of the node is above a threshold, its descriptions are merged, aggregating their attribute values, to enable further matching decisions; if the similarity value of this node has increased, its neighbor nodes are added to the priority queue. This iterative process continues until the priority queue becomes empty. 

\subsubsection{Schema-agnostic methods}

\textit{Collective ER for tree (XML) data} is studied in \cite{DBLP:conf/icde/WeisN06}. Entity descriptions correspond to XML elements composed of text data or other XML elements, and domain experts specify which XML elements are match candidates, thus, initializing a priority queue of comparisons.  Entity dependency takes the following form in this case: an XML element $c$ depends on another XML element $c'$, if $c'$ is a part of the description of $c$. Consequently, identifying the matches of $c$ is not independent of identifying the matches of $c'$. Even if two XML elements are initially considered to be non-matches, they are compared again, if their related elements are marked as matches. A similar approach is based on the intuition that the similarity of two elements reflects the similarity of their data, as well as the similarity of their children \cite{DBLP:conf/iqis/WeisN04}. Following a top-down traversal of XML data, the DELPHI containment metric~\cite{DBLP:conf/vldb/AnanthakrishnaCG02} is used to compare two~elements. 

\vspace{-.1cm}
\begin{example}
Figure~\ref{fig:iterativeERexample}(c) shows two different descriptions of the movie \emph{A Clockwork Orange} in XML. This representation means that the element \emph{movie} consists of the elements \emph{title}, \emph{year} and \emph{cast}, with the last one further consists of \emph{actor} elements. To identify that the two XML descriptions represent the same movie, we can start by examining the cast of the movies. After we identify that actors $a_{11}$ and $a_{21}$ represent the same person, Malcolm McDowell, the chances that the movies $m_1$ and $m_2$ match are increased. They are further increased when we find that actors $a_{12}$ and $a_{22}$ also match, representing Patrick Magee. The same matching process over all the sub-elements of $m_1$ and $m_2$ will finally lead us to identify that $m_1$ and $m_2$ match.
\end{example}

\textit{SiGMa}~\cite{DBLP:conf/kdd/Lacoste-JulienPDKGG13} selects as seed matches the pairs that have identical entity names. Then, it propagates the matching decisions on the compatible neighbors of existing matches. Unique Mapping Clustering is applied for detecting duplicates. For every new matched pair, the similarities of the neighbors are recomputed and their position in the priority queue is updated. 

\textit{LINDA}~\cite{DBLP:conf/cikm/BohmMNW12} follows a very similar approach, which differs from SiGMa mainly in the similarity functions and the lack of a manual relation alignment. LINDA relies on the edit distance of the relation names used in the two KBs to determine if they are equivalent or not. This alignment method makes a strong assumption that descriptions in KBs use meaningful names for relations and similar names for equivalent relations, which is often not true in the Web of Data. Rather than using a similarity threshold, the resolution process in LINDA terminates when the priority queue is empty, or after performing a predetermined number of iterations. 

\textit{RiMOM-IM}~\cite{DBLP:journals/tkde/LiTLL09,DBLP:journals/jcst/ShaoHLWCX16} initially considers as matches entities placed in blocks of size 2. It also uses a heuristic called ``one-left object'':  if two matched descriptions $e_1, e_1'$ are connected via aligned relations $r$ and $r'$ and all their entity neighbors via $r$ and $r'$, except $e_2$ and $e_2'$, have been matched, then $e_2$, $e_2'$ are also considered matches. Similar to SiGMa, RiMOM-IM employs a complex similarity score, which requires the alignment of relations among the KBs.

\textit{PARIS}~\cite{DBLP:journals/pvldb/SuchanekAS11} uses a probabilistic model to identify matching evidence, based on previous matches and the functional nature of entity relations. A relation is considered to be functional if, for a given source entity, there is only one destination entity (e.g., {\small \texttt{wasBornIn}}). The basic matching idea is that if $r(x,y)$ is a function in one KB and $r(x,y')$ is a function in another KB, then $y$ and $y'$ are considered to be matches. The \emph{functionality}, i.e., degree by which a relation is close to being a function, and the alignment of relations along with previous matching decisions determine the decisions in subsequent iterations. The functionality of each relation is computed at the beginning of the algorithm and remains unchanged. Initially, instances with identical values (for all attributes) are considered matches and based on those matches, an alignment of relations takes place. In every iteration, instances are compared based on the newly aligned relations, and this process continues until convergence. In the last step, an alignment of classes (i.e., entity types) also takes place. 

On another line of research, \textit{MinoanER}~\cite{DBLP:conf/edbt/Efthymiou0SC19} executes a non-iterative process that involves four matching rules. First, it identifies matches based on their name (rule \textsf{R1}). This is a very effective and efficient method that can be applied to all descriptions,  regardless of their values or neighbor similarity, by automatically specifying distinctive names of entities based on data statistics. Then, the value similarity is exploited to find matches with many common and infrequent tokens, i.e., strongly similar matches (rule \textsf{R2}). When value similarity is not high, nearly similar matches are identified based on both value and neighbors similarity using a threshold-free rank aggregation function (rule \textsf{R3}). Finally, reciprocal evidence of matching is  exploited  as  a  verification  of  the  returned  results:  only entities  mutually  ranked  in the top matching candidate positions of their unified ranking lists are considered as matches (rule \textsf{R4}). 

\subsection{Learning-based methods}

The first probabilistic model for ER~\cite{fellegi69} used attribute similarities as the dimensions of comparison vectors, each representing the probability that a pair of descriptions match. Following the same conceptual model, a large number of works try to automate the process of learning such probabilities based on manually or automatically generated, or even pre-existing training data. Next, we explore different ways of generating and exploiting training data.

\vspace{4pt}
\noindent\textbf{Supervised Learning.} \textit{Adaptive Matching} \cite{DBLP:conf/kdd/CohenR02} learns from the training data a composite function that combines many attribute similarity measures. Similarly, \textit{MARLIN}~\cite{BilenkoM03} uses labeled data at two levels. 
First, it can utilize trainable string similarity/distance measures, such as learnable edit distance, adapting textual similarity computations to specific attributes. Second, it uses labeled data to train a classifier that distinguishes pairs between matches and non-matches, using textual similarity values for different attributes as features. 

\textit{Gradient-based Matching} \cite{DBLP:journals/dke/Reyes-GalavizPH17} proposes a model that can adjust its structure and parameters based on aggregate similarity scores coming from individual similarity functions on different attributes. Its design allows for locating which similarity functions and attributes are more significant to correctly classify pairs. For its training, it employs a performance index that helps to separate descriptions that have already been matched from those that have not been matched as yet.

\textit{BN-based Collective ER} \cite{DBLP:conf/caise/IoannouNN08} adapts a relationship-based collective ER approach (similar to~\cite{DBLP:conf/sigmod/DongHMN05}) to a supervised learning setting. A Bayesian network is used to capture cause-effect relationships, which are modeled as directed acyclic graphs, and to compute matching probabilities. The lexical similarity in the attribute values of the descriptions as well as their links to existing matches constitute positive matching evidence, which incrementally updates the Bayesian network nodes, similar to the incremental updates that take place in the graph-based dependency model of~\cite{DBLP:conf/sigmod/DongHMN05}.

\textit{GenLink}~\cite{DBLP:journals/pvldb/IseleB12} is a supervised, genetic programming algorithm for learning expressive linkage rules, i.e., functions that assign similarity values to pairs of descriptions. GenLink generates linkage rules that select the important attributes for comparing two descriptions, normalize their attribute values before similarity computations, choose appropriate similarity measures and thresholds, and combine the results of multiple comparisons using linear as well as non-linear aggregation functions. It has been incorporated into the Silk Link Discovery Framework~\cite{DBLP:conf/www/VolzBGK09} (see Section \ref{sec:tools}).

\vspace{4pt}
\noindent\textbf{Weakly Supervised Learning.} Arguably, the biggest limitation of supervised approaches is the need for a labeled dataset, based on which the underlying machine learning algorithm will learn how to classify new instances.
Methods of this category reduce the cost of acquiring such a dataset.

A \emph{transfer learning} approach is proposed in \cite{DBLP:journals/corr/abs-1809-11084} with the aim of adapting and reusing labeled data from a related dataset. The idea is to use a standardized feature space in which the entity embeddings of the reused and the targeted dataset will be transferred. This way, existing labeled data from another dataset can be used to train a classifier that can work with the target dataset, even if there are no explicitly labeled data for the target dataset. A similar transfer learning approach is also followed in~\cite{DBLP:conf/semweb/RongNXWYY12} to infer equivalence links in a linked data setting. 

\textit{Snorkel}~\cite{DBLP:journals/pvldb/RatnerBEFWR17} is a generic tool that can be used to generate training data for a broader range of problems than ER. It relies on user-provided heuristic rules (e.g., several matching functions) to label some user-provided data and evaluate this labeling using a small pre-labeled dataset. Instead of attribute weighting,
Snorkel tries to learn the importance of the provided matching functions. 
This approach of weighting matching rules, instead of features, resembles and complements existing works in ER. For example, the goal in \cite{DBLP:journals/pvldb/WangLYF11} is to identify which similarity measure can maximize a specific objective function for an ER task, given a set of positive and negative examples. Those examples can be generated manually one-by-one, or by leveraging tools like Snorkel. 

\vspace{4pt}
\noindent\textbf{Unsupervised Learning.} \textit{Unsupervised Ensemble Learning} \cite{DBLP:journals/is/JurekHCL17} generates an ensemble of automatic self-learning models that use different similarity measures. To enhance the automatic self-learning process, it incorporates attribute weighting into the automatic seed selection for each of the self-learning models. To ensure that there is high diversity among the selected self-learning models, it utilizes an unsupervised diversity measure. Based on it, the self-learning models with high contribution ratios are kept, while the ones with poor accuracy are discarded.

Rather than relying on domain expertise or manually labeled samples, the unsupervised ER system presented in~\cite{DBLP:journals/ws/KejriwalM15} automatically generates its own heuristic training set. As positive examples are considered the pair of descriptions with very high Jaccard similarity of the token sets in their attribute values. In the context of Clean-Clean ER, having generated the positive example ($e1$, $e2$), where $e1$ belongs to entity collection $\mathcal{E}_1$ and $e2$ to $\mathcal{E}_2$, for every other positive example ($e3$, $e4$), where $e3 \in \mathcal{E}_1$ and $e4 \in \mathcal{E}_2$, it further infers the negative examples ($e1$, $e4$) and ($e3$, $e2$). The resulting training set is first used by the system for Schema Matching to align the attributes in the input datasets. The attribute alignment and the training sets are then used to simultaneously learn two functions, one for Blocking and the other for Matching.

\subsection{Parallel methods}
We now discuss works that are able to leverage massive parallelization frameworks.

A framework for scaling collective ER~\cite{DBLP:journals/tkdd/BhattacharyaG07} to large datasets is proposed in \cite{DBLP:journals/pvldb/RastogiDG11}, assuming a black-box ER algorithm. To achieve high scalability, it runs multiple instances of the ER algorithm in small subsets of the entity descriptions. An initial block collection is constructed based on the similarity of the descriptions using Canopy Clustering~\cite{DBLP:conf/kdd/McCallumNU00}. Each block is then extended by taking its \textit{boundary} with respect to entity relationships. Next, a simple message-passing algorithm is run, to ensure that the match decisions within a block, which might influence the match decisions in other blocks, are propagated to those other blocks. This algorithm retains a list of active blocks, which initially contains all blocks. The black-box ER algorithm is run locally, for each active block, and the newly-identified matches are added in the result set. All the blocks with a description of the newly-identified matches, are set as active. This iterative algorithm terminates when the list of active blocks becomes empty. 

\textit{LINDA}~\cite{DBLP:conf/cikm/BohmMNW12} scales out using MapReduce. The pairs of descriptions are sorted in descending order of similarity and stored in a priority queue. Each cluster node holds: (i) a partition of this priority queue, and (ii) the corresponding part of the entity graph, which contains the descriptions in the local priority queue partition along with their neighbors. The iteration step of the algorithm is that, by default, the first pair in the priority queue is considered to be a match and is then removed from the queue and added to the known matches. This knowledge triggers similarity re-computations, which affect the priority queue by: (i) enlarging it, when the neighbors of the new match are added again to the queue, (ii) re-ordering it, when the neighbors of the identified match move higher in the rank, or (iii) shrinking it, after applying transitivity and the constraint for a unique match per KB. The algorithm stops when the priority queue is empty, or after a specific number of iterations.

Finally, \textit{Minoan-ER} \cite{DBLP:conf/edbt/Efthymiou0SC19} runs on top of Apache Spark. To minimize its overall run-time, it applies Name Blocking, while extracting the top similar neighbors per entity and running Token Blocking. Then, it synchronizes the results of the last two processes: it combines the value similarities computed by Token Blocking with the top neighbors per entity to estimate the neighbor similarities for all entity pairs with neighbors co-occurring in at least one block. Matching rule \textsf{R1} (finding matches based on their name) starts right after Name Blocking, \textsf{R2} (finding strongly similar matches) after \textsf{H1} and Token Blocking, \textsf{R3} (finding nearly similar matches) after \textsf{R2} and the computation of neighbor similarities, while \textsf{R4} (the reciprocity filter) runs last, providing the final, filtered set of matches. During the execution of every rule, each Spark worker contains only the partial information of the blocking graph that is necessary to find the match of a specific node. 

\begin{table}
\vspace{-5pt}
    \caption{Taxonomy of the Matching methods discussed in Section \ref{sec:matching}. MB stands for Merging-based, RB for Relationship-based, S for Supervised, WS for Weakly Supervised and U for Unsupervised Learning.}
    \vspace{-12pt}
	\center \includegraphics[width=0.85\columnwidth]{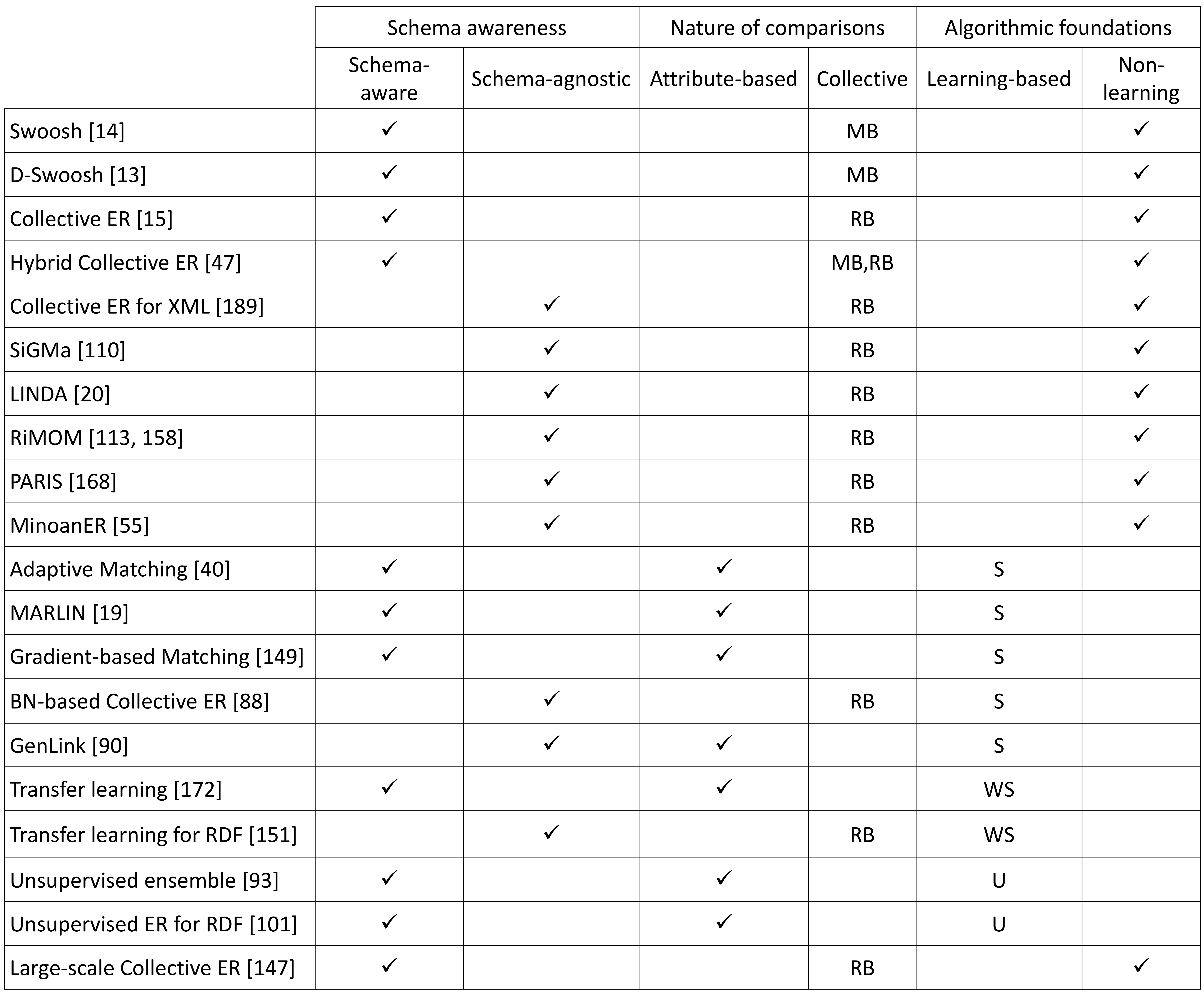}
	\label{tab:matching_taxonomy}
	\vspace{-12pt}
\end{table}

\subsection{Discussion}

Table~\ref{tab:matching_taxonomy} presents an overview of the Matching methods discussed in this section. They are organized according to schema-awareness (schema-aware or schema-agnostic), nature of comparisons (attribute-based or collective), and algorithmic foundations (non-learning or learning-based). Collective methods are further refined as merging-based (MB) or relationship-based (RB),  
and learning-based methods as supervised (S), weakly supervised (WS) and unsupervised (U).

We observe that all schema-agnostic methods that have been proposed are collective, and more specifically, relationship-based. This happens because, unlike the schema-aware methods, the schema-agnostic ones cannot rely on attribute-level similarities for attributes that are not known in advance, or it is not known if they are actually used by the descriptions. Hence, those methods propagate the information provided by entity neighbors as matching evidence whenever possible. 
Consequently, as a rule of thumb that depends on the nature of the input data, we recommend merging-based collective ER methods, which are schema-aware, for data coming from a single dirty entity collection (e.g., for the deduplication of a dirty customer data base) and relationship-based collective ER methods, which are schema-agnostic, for data coming from multiple, curated entity collections (e.g., for finding equivalent descriptions among two or more Web KBs). 

Note that the learning-based methods can be seen as \textit{attribute-based}, since they essentially try to learn the probability that two descriptions match based on previous examples of similar pairs, or \textit{collective}, since their models are trained on sets of pairs, or even on vectorial representations of entity descriptions, or the words used in the values of those descriptions. For completeness, Table \ref{tab:matching_taxonomy} classifies them as attribute-based, following the traditional learning approach, because their collective nature cannot be easily labeled as merging-based or relationship-based. 
We believe that the learning-based methods are gaining ground as new and more effective ways to represent individual or groups of entity descriptions appear (see Section~\ref{sec:deepLearning}). The emergence of weakly supervised and transfer-learning methods seem to alleviate the problem of generating a labeled set for training data. Therefore, when labeled examples are available (e.g., in transfer learning), or are easy to generate using existing tools (e.g.,~\cite{DBLP:journals/pvldb/RatnerBEFWR17}), and the test data are not expected to deviate considerably from the training data, those methods seem to be the most promising ones. Before choosing learning-based or non-learning methods, one should also consider the desired frequency of re-training a new classification model, the memory footprint of each method (i.e., whether the whole model needs to reside in memory or not) and the time needed for training and classification. 

In general, recent studies \cite{DBLP:journals/corr/abs-1710-00597,DBLP:journals/pvldb/KondaDCDABLPZNP16,DBLP:conf/sigmod/MudgalLRDPKDAR18} show that the learning-based techniques achieve higher accuracy than the rule-based ones that are used in several practical scenarios. Yet, despite some past efforts (e.g.,~\cite{DBLP:journals/jodsn/IoannouRV13,DBLP:journals/dke/KopckeR10,DBLP:journals/pvldb/KopckeTR10}), we notice the lack of a systematic benchmarking of matching methods. A comprehensive benchmark should evaluate effectiveness (i.e., quality of the output matches), time and space efficiency (i.e., the time required for pre-processing, training, and matching, the memory and disk space required by each method), and scalability (i.e., using the same computational and storage resources, what is the data limit that each method can handle).

\section{Clustering Methods}
\label{sec:clustering}

Typically, clustering constitutes the final task in the end-to-end ER workflow, following Matching. Its input comprises the \textit{similarity graph}, where the nodes correspond to the descriptions and each edge connects a pair of descriptions that were compared during Matching; the edge weights, typically in $[0,1]$, are analogous to the matching likelihood of the adjacent descriptions.
Clustering aims to infer more edges from indirect matching relations, while discarding edges that are unlikely to connect duplicates in favor of edges with higher weights. Hence, its end result is a set of \textit{entity clusters}, each of which comprises all descriptions that correspond to the same, distinct~real-world~object. 

In the simplest case, \textit{Connected Components} \cite{DBLP:journals/pvldb/HassanzadehCML09,DBLP:journals/csimq/SaeediNPR18} is applied to compute the transitive closure of the detected matches. This naive approach increases recall, but is rather sensitive to noise. False positives have a significant impact on precision, leading to entity clusters that are dominated by non-matching descriptions. For this reason, more advanced clustering techniques have been proposed to leverage the weighted edges in the similarity graph. In general, these techniques are distinguished into three categories, according to the type of the ER task at hand:

1) For Clean-Clean ER, clustering typically relies on the 1-1 correspondence between the input data sources. The most popular technique is \textit{Unique Mapping Clustering} \cite{DBLP:conf/kdd/Lacoste-JulienPDKGG13,DBLP:conf/cikm/BohmMNW12}, which first sorts all edges in decreasing weight. At each iteration, the top edge is considered a match, if none of the adjacent descriptions has already been matched. The process ends when the top edge has a similarity lower than a threshold $t$. Essentially, this approach provides an efficient solution to the \textit{Stable Marriage} problem for unequal sets \cite{mcvitie1970stable}, given that Clean-Clean ER forms a (usually unbalanced) bipartite similarity graph. The \textit{Hungarian algorithm} is also applicable, though at a much higher computational cost,
unless an approximation is used, as in \cite{DBLP:journals/eor/DiazF01,DBLP:journals/jacm/Kurtzberg62}.

2) For Dirty ER, the core characteristic of clustering algorithms is that they produce a set of disjoint entity clusters without requiring as input the number of clusters or any labeled dataset for training \cite{DBLP:journals/pvldb/HassanzadehCML09}. \textit{Center Clustering} \cite{DBLP:conf/webdb/HaveliwalaGI00} iterates once over all edges and creates clusters around nodes that are selected as centers. Its functionality is enhanced by \textit{Merge-Center Clustering} \cite{DBLP:journals/vldb/HassanzadehM09}, which merges together clusters with centers similar to the same node. \textit{Star Clustering} \cite{DBLP:journals/jgaa/AslamPR04} begins with sorting all similarity graph nodes in descending order of degree. Then, the top node becomes the center of a cluster that includes all its direct neighbors. The same process is repeatedly applied to the remaining nodes, until all nodes belong to a cluster. The resulting clusters are overlapping, unless post-processing assigns each node to a single cluster.  \textit{Ricochet Clustering} \cite{DBLP:conf/dasfaa/WijayaB09} comprises a family of techniques based on two alternating stages: the first one determines the centers of clusters (like Star Clustering), while the second one (re-)assigns nodes to cluster centers (like~K-Means).

Other techniques focus on the relative strength of the links inside and across clusters, i.e., the \textit{intra-} and \textit{inter-cluster} edges. \textit{Markov Clustering} \cite{van2000graph} uses random walks to strengthen the intra-cluster edges, while weakening the inter-cluster ones. \textit{Cut clustering} \cite{DBLP:journals/im/FlakeTT03} iteratively identifies the minimum cut of maximum flow paths from a node to an artificial sink node. This way, it detects small inter-cluster cuts, while strengthening intra-cluster links. \textit{Correlation Clustering} \cite{DBLP:journals/ml/BansalBC04} solves an optimization task, where the goal is to maximize the sum of the intra-cluster edges, while minimizing the sum of the inter-cluster ones. This is an NP-hard problem that is typically solved through approximations, such as \textit{Clustering Aggregation} \cite{DBLP:journals/tkdd/GionisMT07} and \textit{Restricted Correlation Clustering} \cite{rcclustering}. The latter is a semi-supervised approach that leverages a small labeled dataset, which is carefully selected via an efficient sampling procedure based on LSH. 

3) For Multi-source ER \cite{DBLP:journals/csimq/SaeediNPR18}, we can use most algorithms for Dirty ER, but the multitude of input entity collections calls for specialized clustering methods. \textit{SplitMerge} \cite{DBLP:conf/icdm/NentwigGR16} applies Connected Components clustering and cleans the resulting clusters by iteratively removing entities with low similarity to other cluster members. Then, it merges similar clusters that are likely to correspond to the same real-world entity. \textit{CLIP} \cite{DBLP:conf/esws/SaeediPR18} assumes duplicate-free entity collections as input. First, it computes the transitive closure of the strong links, i.e., the edges that correspond to the maximum weight per source (entity collection) for both adjacent nodes. The remaining graph is cleaned from the weak links, i.e., the edges that do not correspond to the maximum weight per source for neither adjacent node. Finally, the transitive closure is computed and its clusters are processed to ensure that they contain at most one description per source. 

\vspace{3pt}
\noindent
\textbf{Discussion.} The relative performance of Dirty ER methods has been experimentally evaluated in \cite{DBLP:journals/pvldb/HassanzadehCML09}. As expected, Connected Components exhibits the worst accuracy. Ricochet Clustering performs well only over entity collections with uniformly distributed duplicates, while Markov Clustering consistently achieves top performance. Surprisingly enough, the highly scalable, single-pass algorithms Center and Merge-Center clustering provide comparable, if not better, results than more complex techniques, like Cut and Correlation Clustering. 

The relative performance of Multi-source ER algorithms is examined in \cite{DBLP:journals/csimq/SaeediNPR18,DBLP:conf/adbis/SaeediPR17}, using their parallelization in Apache Flink. The experiments show that SplitMerge and CLIP achieve the top performance, with the latter providing a better balance between effectiveness and time efficiency.

\section{Budget-aware ER}
\label{sec:progressiveER}

Unlike the budget-agnostic methods presented above, budget-aware ER provides the best possible \textit{partial solution}, when the response time or the available computational resources are constrained. It is driven by a \textit{pay-as-you-go} paradigm that sacrifices the completeness of results, when the number of data sources or the amount of data to be processed is ever increasing. For example, the number of high-quality HTML tables on the Web is in the hundreds of millions, while the Google search system alone has indexed $\sim$26 billion datasets \cite{DBLP:conf/pods/GolshanHMT17}. This unprecedented volume of data can only be resolved progressively, using matching pairs from former iterations to generate more accurate candidate pairs in the latter iterations as long as the allocated budget is not exhausted.

Typically, budget-aware methods rely on blocking as a pre-processing task that identifies similar entity descriptions. They differ, though, on how they leverage the resulting blocks in the Planning step - see Figure \ref{fig:workflow}(b).
Four categories of granularity functionality are defined \cite{simonini2018schema}: 
\begin{enumerate}
    \item \textit{Block-centric methods} produce a list of blocks that are sorted in descending order of the likelihood that they include duplicates among their descriptions. All the comparisons inside each block are generated iteratively, one block at a time, following that ordered list.
    \item \textit{Comparison-centric methods} provide a list of description pairs sorted in descending order of matching likelihood. These pairs of descriptions are emitted iteratively, one at a time, following that ordered list.
    \item \textit{Entity-centric methods} provide a list of descriptions sorted in descending order of duplication likelihood. All comparisons of every description are generated iteratively, one description at a time, following that ordered list.
    \item The \textit{hybrid methods} combine characteristics from two or all of the previous categories.
\end{enumerate}

Depending on their blocking keys, budget-aware methods are further classified into \cite{simonini2018schema}: 
\begin{enumerate}
    \item \textit{Sort-based methods}, which rely on the similarity of blocking keys. They produce a list of descriptions by sorting them alphabetically, according to their blocking keys, and assume that the matching likelihood of two descriptions is analogous to their proximity after sorting.
    \item \textit{Hash-based methods}, which consider identical blocking keys and typically assume redundancy-positive blocks, i.e., the similarity of two descriptions is proportional to their common blocks.
\end{enumerate}

In the sequel, we examine separately the schema-aware and the schema-agnostic methods.

\subsection{Schema-aware methods}

The budget-aware methods that are suitable for structured data rely on schema knowledge. This means that their performance depends heavily on the attribute(s) that provide the schema-aware blocking keys they leverage, typically requiring domain experts to fine-tune them.

The core comparison-centric method is \textit{Progressive Sorted Neighborhood} (\textbf{PSN}) \cite{DBLP:journals/tkde/WhangMG13}. Based on Sorted Neighborhood \cite{DBLP:conf/sigmod/HernandezS95}, it associates every description with a schema-aware blocking key. Then, it produces a \textit{sorted list of descriptions} by ordering all blocking keys alphabetically. Comparisons are progressively defined through a sliding window, $w$, whose size is \textit{iteratively incremented}: initially, all descriptions in consecutive positions ($w$=1) are compared, starting from the top of the list; then, all descriptions at distance $w$=2 are compared and so on, until termination. 

The above approach produces a \textit{static} list of comparisons, which remains immutable, regardless of the duplicates that are identified. As a result, PSN cannot react to the skewed distribution of duplicates. To ameliorate this issue, a \textit{dynamic} version of the algorithm was proposed in \cite{DBLP:journals/tkde/PapenbrockHN15}. Its functionality is integrated with Matching to adjust the processing order of comparisons on-the-fly. Arranging the sorted descriptions in a two-dimensional array $A$, if position $A(i,j)$ corresponds to a duplicate, the processing moves on to check positions $A(i+1,j)$ and $A(i,j+1)$.

The same principle lies at the core of the dynamic, block-centric method \textit{Progressive Blocking}~\cite{DBLP:journals/tkde/PapenbrockHN15}. Initially, a set of blocks is created and its elements are arranged in a two-dimensional array $A$. Then, all comparisons are executed inside every block, measuring the number of duplicates per block. Starting from the block with the highest density of duplicates in position $A(i,j)$, its descriptions are compared with those in the blocks $A(i+1,j)$ and $A(i,j+1)$ in order to identify more~matches.

A static, block-centric method is the \textit{Hierarchy of Record Partitions} (\textbf{HRP}) \cite{DBLP:journals/tkde/WhangMG13}, which  presumes that the distance of two records can be naturally estimated through a certain attribute (e.g., product price). Essentially, it builds a hierarchy of blocks, such that the matching likelihood of two descriptions is proportional to the level in which they co-occur for the first time: the blocks at the bottom of the hierarchy contain the descriptions with the highest matching likelihood, and vice versa for the top hierarchy levels. Then, the hierarchy of blocks is progressively resolved, level by level, from the leaves to the root. A variation of this approach is presented in \cite{DBLP:conf/icde/AltowimM17}: every block is divided into a hierarchy of child blocks and an advanced strategy optimizes their processing on MapReduce.

An entity-centric improvement of the HRP is the \textit{Ordered List of Records} \cite{DBLP:journals/tkde/WhangMG13}, which converts the hierarchy of blocks into a list of records sorted by their likelihood to produce matches. In this way, it trades lower memory consumption for a slightly worse performance than HRP.

Finally, a progressive approach for Multi-source ER over different entity types is proposed in \cite{DBLP:journals/pvldb/AltowimKM14}. During the scheduling phase, it divides the total cost budget into several windows of equal cost. For each window, a comparison schedule is generated by choosing the one with the highest expected benefit among those with a cost lower than the current window. The cost of a schedule is computed by considering the cost of finding the description pairs and the cost of resolving them. Its benefit is determined by how many matches are expected to be found by this schedule and how useful they will be to identify more matches within the cost budget. After a schedule is executed, the  matching decisions are propagated to all related comparisons so that they are more likely to be chosen by the next schedule. The algorithm terminates upon reaching the cost budget.

\vspace{-5pt}
\subsection{Schema-agnostic methods}

The budget-aware methods for semi-structured data rely on an inherently schema-agnostic functionality that completely disregards any schema information. Thus, they are independent of expert knowledge and require no labeled data for learning how to rank comparisons, blocks or descriptions.

The cornerstone of sort-based methods is the \textit{Neighbor List} \cite{simonini2018schema}, which is created by the schema-agnostic adaptation of Sorted Neighborhood \cite{DBLP:journals/pvldb/0001APK15}: every token in any attribute value is considered as a blocking key and all descriptions are sorted alphabetically according to these keys. Thus, each description appears in the Neighbor List as many times as the number of its distinct tokens. 

The naive progressive approach would be to slide a window of increasing size along this list, incrementally executing the comparisons it defines, as in PSN. This approach, however, results in many repeated comparisons and a random ordering of descriptions with identical keys.

To ameliorate this issue, \textit{Local Schema-agnostic PSN} \cite{simonini2018schema} uses weights based on the assumption that the closer the blocking keys of two descriptions are in the Neighbor List, the more likely they are to be matching. Every comparison defined by the current window size is associated with a numerical estimation of the likelihood that it involves a pair of matches through the schema-agnostic weighting function $\frac{fr_{j,i}}{fr_i+fr_j-fr_{i,j}}$, where $fr_k$ is the number of blocking keys associated with description $e_k$ (i.e., its occurrences in the Neighbor List), while $fr_{j,i}$ denotes the frequency of comparison $<e_i, e_j>$ within the current window. All repeated comparisons within every window are eliminated, but there is no way to avoid emitting the same comparison in other window sizes. To address this drawback, \textit{Global Schema-agnostic PSN} \cite{simonini2018schema} defines a global execution order for all comparisons in a specific range of window sizes $[1, w_{max}]$, using the same weighting function.

A different approach is implemented by the hash-based method \textit{Progressive Block Scheduling} \cite{simonini2018schema}. First, the input blocks are ordered in increasing cardinality such that the fewer comparisons a block entails, the higher it is ranked. Then, the sorted list of blocks is processed, starting from the top-ranked (i.e., smallest) block. Inside every block, one of Meta-blocking's weighting schemes is used to specify the processing order of comparisons, from the highest weighted to the lowest one. During this process, all repeated comparisons are discarded before computing their weight. 

Finally, \textit{Progressive Profile Scheduling} \cite{simonini2018schema} is a hybrid method that relies on the notion of \textit{duplication likelihood}, i.e., the likelihood of an individual description to have one or more matches. This is estimated as the average edge weight of its node in the corresponding blocking graph. This method processes the input descriptions in decreasing duplication likelihood. For each description, all non-repeated comparisons that entail it are ordered in decreasing weight, as estimated through a Meta-blocking weighting scheme, and the top-k ones are emitted. 

\subsection{Discussion}

All budget-aware methods apply ER in a pay-as-you go manner. To address Volume, they all rely on blocking methods. The schema-agnostic budget-aware methods are also capable of addressing Variety. Table \ref{fig:progressiveTaxonomy} organizes all methods discussed above into a taxonomy formed by the four aforementioned criteria: schema-awareness, functionality of blocking keys, granularity of functionality and type of ordering. We observe that there is no dynamic schema-agnostic method that adapts its processing order as more duplicates are identified. More research is required towards this direction. For dynamic schema-aware methods, a noisy matching method should be used, instead of the ideal one that is currently considered. Intelligent ways for tackling the errors introduced by noisy matchers are indispensable for a realistic budget-aware scenario.

\begin{table*}[t]\centering
\vspace{-5pt}
\caption{A taxonomy of the budget-aware methods discussed in Section \ref{sec:progressiveER} (in the order of presentation).}
\vspace{-5pt}
\includegraphics[width=0.99\linewidth]{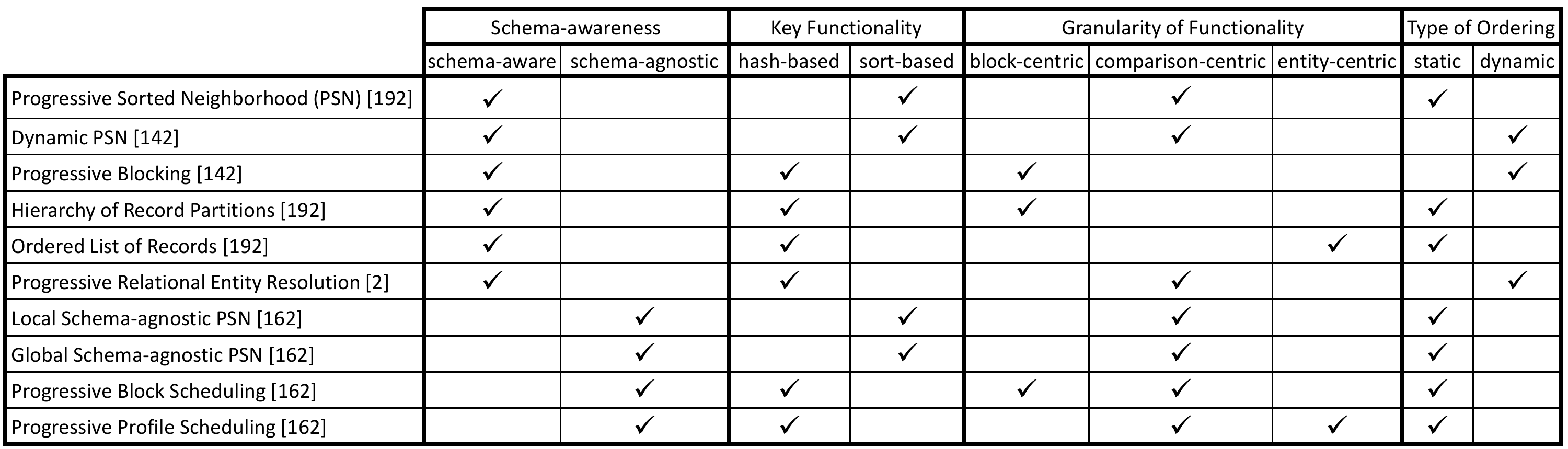}
\label{fig:progressiveTaxonomy}
\vspace{-5pt}
\end{table*}

Regarding the relative performance of static methods, the schema-agnostic ones consistently outperform the schema-aware ones over several established structured datasets \cite{simonini2018schema}. Among the schema-agnostic methods, the two sort-based ones achieve the best performance for structured datasets, with the difference between them being statistically insignificant. As a result, Local PSN is more suitable in cases of limited memory, but all other settings call for Global PSN, given that it avoids multiple emissions of the same comparisons. For large, heterogeneous datasets, Progressive Profile Scheduling exhibits the overall best performance, followed by Progressive Block Scheduling.

\section{Incremental ER}
\label{sec:incrementalER}

Some Big Data applications need to resolve descriptions that arrive in high Velocity streams or are provided as queries against a known entity collection. Rather than a static, offline process over all available entity descriptions, such applications process as much entities as needed as long as they resolve specific (query) descriptions in (near) real time. The same applies to clean, but evolving data repositories, such as data warehouses and knowledge bases, where new entities should be incrementally added, without repeating the entire ER process to the already matched~descriptions.

As an example, consider an application resolving the entities described across news feeds, which arrive in a streaming fashion \cite{DBLP:conf/edbt/KarapiperisGV18,DBLP:conf/sac/AraujoSPNN20,DBLP:conf/icdm/BilenkoBS05}. A journalist using this application could be provided with several facts regarding a breaking news story (e.g., persons, buildings, services affected by an earthquake), as they get published by different agencies or witnesses, enabling her/him to form a complete picture of the events as they occur, in real-time. This would require storing only some parts of the entire entity collection, and discarding the rest, as more descriptions are fed to the system. To evaluate which parts of the collection are more useful to keep, we can design different strategies. For example, we may want to keep the latest entities, since new input entities are more likely to be connected to them.
Another strategy would be to keep the entities with many relationships with other entities, since they are more likely to influence the matching decisions.

Such applications call for small memory footprint and low latency, rendering inapplicable the \textit{static} approaches described above. 
Novel techniques that \textit{dynamically} adapt to data are required.
Note that we could distinguish the dynamic methods into those answering to a user-provided query and those resolving streams of entities, but this distinction is orthogonal - streaming methods can be seen as query-based ones that handle streams of queries instead of a single query (e.g., \cite{DBLP:conf/edbt/KarapiperisGV18}).

\subsection{Dynamic Blocking}

Unlike the works in Section~\ref{sec:blocking}, which produce immutable (static) blocks, the dynamic indexing techniques update their blocks, depending on the descriptions that are submitted as~queries.

One of the earliest approaches is the \textit{Similarity-aware Index}~\cite{christenDI}. The main idea is to pre-calculate similarities between the attribute values that co-occur in blocks in order to avoid similarity calculations at query time, and minimizing response time. This approach uses three indexes that associate blocking keys to attribute values, that contain pre-calculated similarities between attribute values that co-occur in a block, and that associate distinct attribute values with record ids. 

This approach is extended by \textit{DySimII}~\cite{10.1007/978-3-642-40319-4_5} so that all three indexes are updated as query entities arrive. Both its average record insertion time and its average query time remain practically stable, even when the index size grows. Interestingly, the index size can be reduced, without any significant loss in recall, by indexing only a certain portion of the most frequent attribute values.

On another line of research, \textit{F-DySNI}~\cite{DBLP:conf/cikm/RamadanC14,Ramadan:2015:DSN:2836847.2816821} extends the Sorted Neighborhood method by converting the sorted list of blocking keys into an index tree that is faster to search. This is actually a braided AVL tree, i.e., a combination of a height balanced binary tree and a double-linked list \cite{rice2007braided}: every tree node is linked to its alphabetically sorted predecessor node, to its successor node and to the list of ids of all entities that correspond to its blocking key. F-DySNI actually employs a forest of such index trees, with each tree associated with a different blocking key definition. This forest is updated whenever a query entity arrives and is compatible with both a fixed and an adaptive window. The former defines the rigid number of neighboring nodes that are considered, while the latter considers only the neighbors that exceed a predetermined similarity threshold.

Finally, summarization algorithms for speeding up dynamic ER are presented in \cite{DBLP:conf/edbt/KarapiperisGV18}. \textit{SkipBloom} summarizes the input descriptions, using their blocking keys, to accelerate comparisons. \textit{BlockSketch} summarizes a block to achieve a fixed number of comparisons per given entity description during Matching, yielding a bounded computational time. Each block is split into sub-blocks based on the distances of the block contents to the blocking key. Each query description is then compared against the sub-block with the smallest distance. to its contents \textit{SBlockSketch} adapts BlockSketch to streaming data, maintaining a fixed number of blocks in memory, with a time overhead each time any of those blocks needs to be replaced with blocks residing in secondary storage. To minimize this overhead, a selection algorithm chooses the blocks to be replaced (considering age and size).

\subsection{Dynamic Matching}

These methods resolve online parts of the entity collection that are of interest to a user/application.

\textit{Query-driven ER} \cite{DBLP:journals/jair/BhattacharyaG07} uses a two-stage expand-and-resolve query processing strategy. First, it extracts the related descriptions for a query using two expansion operators. Then, it resolves the extracted descriptions collectively, leveraging an existing relevant technique \cite{DBLP:journals/tkdd/BhattacharyaG07}. Due to~the~complexity of the collective ER strategy, this approach cannot provide real-time answers for large~datasets.

In \textit{Query-driven ER with uncertainty} \cite{DBLP:journals/pvldb/IoannouNNV10}, the attribute-level facts for the input entities are associated with a degree of uncertainty, reflecting the noise from imperfect extraction tools. Matches are identified using existing ER algorithms and are assigned a probability value. At this offline stage, no merging takes place. When a query arrives, the descriptions that need to be merged in order to provide an answer to the query are identified. Then, different merging scenarios are explored and the one with minimum uncertainty is selected and returned~as~an~answer. 

\textit{UDD}~\cite{DBLP:journals/tkde/SuWL10} is an unsupervised
method that identifies matches from the results of a query over multiple Web KBs. First, it removes duplicate descriptions stemming from the same KB, and it generates a training set. 
Based on this set of non-matching examples, as well as on similarity computations between descriptions, it iteratively identifies matches in the query results 
through two cooperating classifiers: a weighted component similarity summing and an SVM.

\textit{Sample-and-clean} \cite{DBLP:conf/sigmod/WangKFGKM14} leverages sampling to improve the quality of aggregate numerical queries on large datasets that are too expensive to resolve online. It resolves a small data sample and exploits those results to reduce the impact of duplicates on the approximate answers to~aggregate~queries. 

\textit{QuERy}~\cite{DBLP:journals/pvldb/AltwaijryMK15} aims to answer join queries over multiple, overlapping data sources, operating on a block level. It identifies which blocks need to be resolved for the requested join and then assumes that any matching method can be applied for the matching task. 

Complementary to this work, \textit{QDA}~\cite{DBLP:journals/tkde/AltwaijryKM17} tries to reduce the data cleaning overhead and issues the minimum number of necessary steps to answer SQL-like selection queries that do not involve joins, in an entity-pair level. It performs vestigiality analysis on each block individually to identify matching decisions whose answers are guaranteed to not affect the query answers and, thus, need not be performed, reducing the matching tasks. In fact, it creates an entity graph for the contents of a block and resolves edges belonging to cliques that may affect the query answer. As opposed to Sample-and-Clean~\cite{DBLP:conf/sigmod/WangKFGKM14}, QDA provides exact query~results. 

Finally, \textit{Adaptive Product Normalization}~\cite{DBLP:conf/icdm/BilenkoBS05} presents an online supervised learning approach for resolving different descriptions of the same product. 
The steps of this approach include: (i) blocking~\cite{DBLP:conf/kdd/McCallumNU00}, which defines an initial set of basis functions to compute the similarity between specific attributes of the descriptions, (ii) a learning algorithm for training the parameters of a composite similarity function, and (iii) clustering~\cite{DBLP:journals/csur/JainMF99}. The composite similarity function is trained incrementally, using an efficient, online variation of the voted perceptron algorithm~\cite{DBLP:journals/ml/FreundS99}. 

\subsection{Dynamic Clustering}

Special care should be taken to update the entity clusters in an efficient way, as more entities arrive in the form of queries or streams. To this end, \textit{Incremental Correlation Clustering} \cite{Gruenheid:2014:IRL:2732939.2732943} supports all kinds of updates (i.e., inserting, deleting and changing individual descriptions from clusters as well as merging and splitting entire clusters), without requiring any prior knowledge of the number of clusters. It also allows for fixing prior errors in view of new evidence. Due to its high complexity, though, a greedy approximation of polynomial time is also proposed. Constrained versions of incremental correlation clustering in other contexts have been proposed in \cite{DBLP:journals/siamcomp/CharikarCFM04,DBLP:conf/stacs/MathieuSS10}.

\begin{table}[t]\centering
\vspace{-5pt}
\caption{A taxonomy of the incremental methods discussed in Section \ref{sec:incrementalER} (in the order of presentation).}
\vspace{-8pt}
\includegraphics[width=0.99\linewidth]{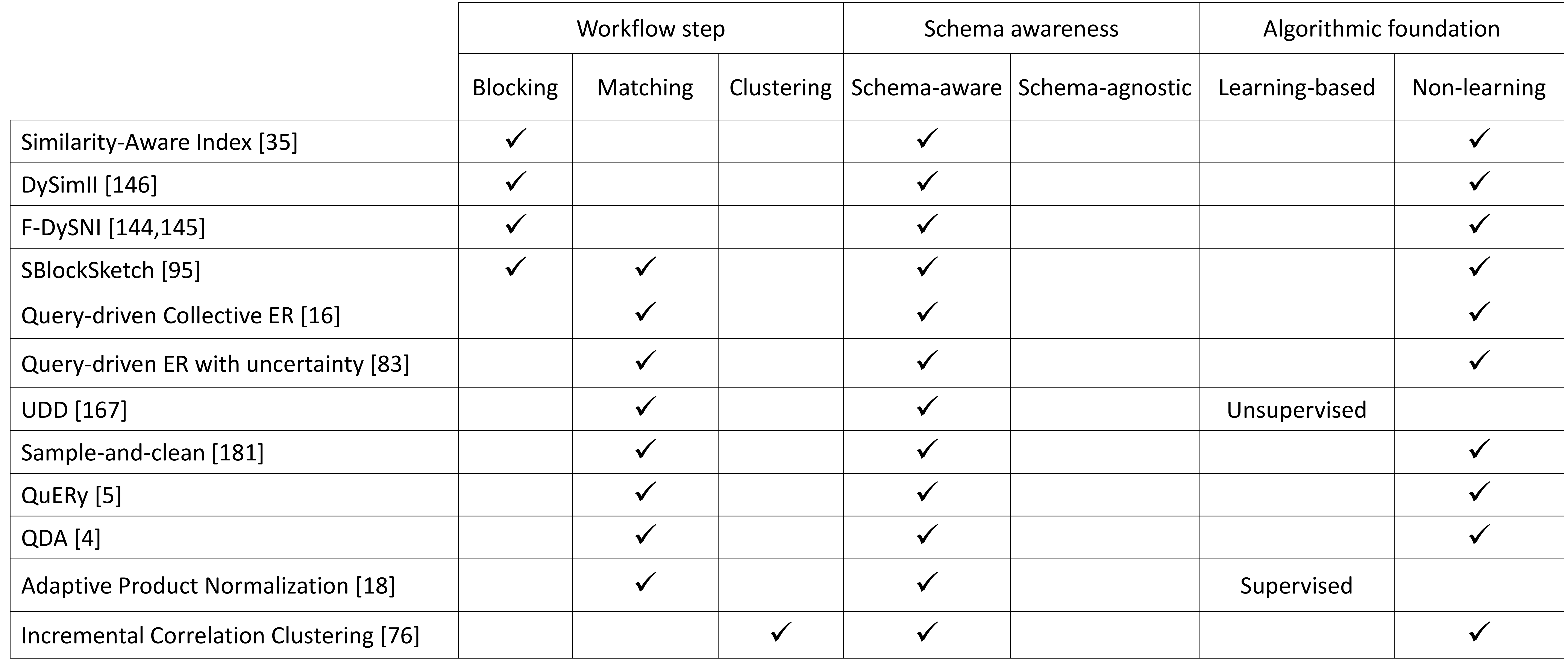}
\label{fig:incrementalTaxonomy}
\vspace{-10pt}
\end{table}

\subsection{Discussion}

Table \ref{fig:incrementalTaxonomy} organizes all methods discussed in this section into a taxonomy formed by three criteria: the ER workflow task corresponding to each method, its schema-awareness and its algorithmic foundation (learning-based or non-learning). These works are crafted for resolving entities in (near) real time, not necessarily covering the whole input entity collections, but only a subset that is associated with a user-defined query or a stream of descriptions. In these cases, resolving the whole input set of descriptions would be unnecessarily costly in terms of time and resources. We believe that in the new Big Data era of unprecedented Volume and Velocity, incremental ER methods are  becoming far more prevalent, gradually displacing traditional, batch ER methods. Yet, all existing methods are schema-aware, being incapable of addressing Variety. More research is required towards schema-agnostic methods or other approaches that inherently support Variety. This also requires the development of incremental schema-agnostic block processing techniques.

\section{Other ER Methods}
\label{sec:other}
We now cover important ER systems and methods complementary to those presented above.

\subsection{Deep Learning} 
\label{sec:deepLearning}

The latest developments in deep learning have greatly influenced research in ER. The basic constructs of deep learning methods for ER are Recurrent Neural Networks (\textit{RNNs})~\cite{DBLP:journals/neco/WilliamsZ89,DBLP:journals/cogsci/Elman90} and word embeddings~\cite{DBLP:journals/jmlr/BengioDVJ03}. 
RNNs are neural networks with a dynamic temporal behavior. The neurons are fed information not only from the previous layer, but also from their own previous state in time, to process sequences of inputs.  
Word embeddings are vectorial representations of words, enabling words or phrases to be compared using their vectors. Word embeddings are commonly used with RNNs for speech recognition~\cite{DBLP:conf/interspeech/MesnilHDB13} and similar NLP tasks~\cite{DBLP:conf/emnlp/ChoMGBBSB14}.

\textit{AutoBlock}~\cite{zhang2019autoblock} trains on a set of matches to perform Blocking. First, it converts every token in an attribute value into a word embedding. Then, a neural network combines word embeddings into several attribute embeddings per description, which are fed into multiple indexing functions. 
The blocking model is learned from training data so that the difference between matching and non-matching descriptions is maximized.
LSH is used to detect the most likely matches per description.

\textit{DeepER}~\cite{DBLP:journals/corr/abs-1710-00597} explores two methods to generate entity embeddings, i.e., vectorial representations of entity descriptions. The first one exploits word embeddings of tokens appearing in the values of the descriptions, while the latter uses RNNs to convert each description to a vector. 
\textit{DeepER} can operate both with pre-trained word embeddings~\cite{DBLP:conf/emnlp/PenningtonSM14}, and without, proposing ways to create and tune such embeddings, customized for ER. The embedding vector of every description is indexed by LSH, whose parameters are set according to a theoretical analysis and the desired performance. 
Then, each entity creates a block that contains its top-N nearest neighbors.
We note that more efficient high-dimensional vector similarity methods (than LSH) are now available~\cite{DBLP:journals/pvldb/EchihabiZPB19}.

\textit{DeepMatcher}~\cite{DBLP:conf/sigmod/MudgalLRDPKDAR18} extends \textit{DeepER} by introducing an architecture template for deep learning ER methods with three main modules: (i) attribute embedding, which converts sequences of words used in the attribute values of an entity description to word embedding vectors; (ii) attribute similarity representation, which applies a similarity function on the attribute embeddings of two descriptions to obtain a final similarity value of those descriptions (i.e., it learns the similarity function); and (iii) a classifier, which uses the similarities between descriptions as features for a classifier that decides if a pair of descriptions is a match (i.e., it learns the match function). 
For each module, several options are available. 
The main ones (e.g., character-level vs word-level embeddings, pre-trained vs learned embeddings, fixed vs learnable similarity function) are used as representative points for those modules and are experimentally evaluated, showing their strengths and weaknesses.

\textit{Multi-Perspective Matching}~\cite{DBLP:conf/ijcai/FuHSCZWK19} adaptively selects (among the similarity measures of \textit{DeepMatcher}'s RNN, the Hybrid similarities for textual attributes, and several established approaches for string and numeric attributes) the optimal similarity measures for heterogenous attributes. First, a unified model for all attributes is built and the supported similarity measures are applied to every attribute value pair. 
A gate mechanism adaptively selects the most appropriate similarity measure per attribute and the selected measures are concatenated into a comparison vector. 
Finally, a neural network receives the comparison vector as input and produces the matching probability as output.

Other works examine ways of optimizing the use of Deep Learning techniques: to minimize the number of required labelled instances, transfer learning is examined in \cite{DBLP:conf/www/ZhaoH19} and pre-trained subword embeddings are combined with transfer and active learning in \cite{DBLP:conf/acl/KasaiQGLP19}; the use of the main attention-based transformer architectures is examined in \cite{DBLP:conf/edbt/BrunnerS20}; pre-trained word embeddings are coupled with online user reviews for each entity description (e.g., restaurant) in \cite{DBLP:conf/www/SchneiderMD18}.

As we have seen, conventional ER methods identify similar entities based on symbolic features (e.g., names, textual descriptions and attribute values). 
However, the computation of feature similarity often suffers from the semantic heterogeneity between different Knowledge Graphs (KG). 
Recently, representation learning techniques have been proposed for Clean-Clean ER, also called \textit{Entity Alignment}, where the key idea is to learn embeddings of KGs, such that entities with similar neighbor structures in the KG have a close representation in the embedding space. 
While several existing techniques learn entity embeddings 
in the context of the same KG, doing the same for entities of different KGs 
remains an open challenge. 
In this setting, \textit{MTransE}~\cite{DBLP:conf/ijcai/ChenTYZ17} learns a mapping between two 
KG embedding spaces, using a seed set of aligned entities from the two KGs, 
though, this is rarely available. 
\textit{JAPE}~\cite{DBLP:conf/semweb/SunHL17} jointly trains the attribute and structure embeddings using skip-gram and translational models, respectively, to align entities. 
\textit{GCN-Align}~\cite{DBLP:conf/emnlp/WangLLZ18} employs Graph Convolutional Networks (GCNs) to model entities based on their neighborhood information. However, GCN-Align only considers the equivalent relations between entities, neglecting the use of additional KG relationships. \textit{IPTransE}~\cite{DBLP:conf/ijcai/ZhuXLS17} and \textit{BootEA}~\cite{DBLP:conf/ijcai/SunHZQ18} integrate knowledge among different KGs by enlarging the training data (prior alignments) in a bootstrapping way. 
\textit{KDCoE}~\cite{DBLP:conf/ijcai/ChenTCSZ18} iteratively co-trains multilingual KG embeddings and fuses them with entity description information for alignment. 
The above iterative methods improve performance mainly by increasing the number of pre-aligned training entity pairs, a strategy 
that could benefit most alignment approaches. 
Non-iterative methods could achieve better results through bootstrapping.

Methods leveraging additional types of features to refine relation-based embeddings include the following.
\textit{AttrE}~\cite{DBLP:conf/aaai/TrisedyaQZ19} uses character-level literal embeddings over a unified vector space for the relationship embeddings after merging the two KGs based on predicate similarity (i.e., predicate alignment).
\cite{DBLP:conf/ijcai/ZhangSHCGQ19} introduces a framework that unifies multiple views of entities to learn embeddings for entity alignment that is capable of incorporating new features. 
Specifically, it embeds entities based on the views of entity names, relations and attributes, with several combination strategies, and considers cross-KG inference methods to enhance the alignment between two KGs. 
A thorough experimental evaluation of supervised and semi-supervised methods for embedding-based entity alignment has been conducted in~\cite{DBLP:journals/corr/abs-2003-07743}. The results on sparse and dense datasets recognize the difficulty of existing methods in aligning (the many) long-tail entities~\cite{DBLP:journals/pvldb/LiDLL17}. Finally, we note that the hierarchical structure of KGs (in particular, ontologies) has not been well studied in this context. Thus, more complex KG embeddings (going beyond Euclidean models) are worth exploiting~\cite{DBLP:conf/nips/NickelK17}.

\subsection{Crowdsourcing-based ER methods}

\textit{Crowd-sourcing} is a recent discipline that examines ways of pushing difficult tasks, called \textit{Human Intelligence Tasks} (\textsf{HITs}), to humans, a.k.a., \textit{workers}, at a small price~\cite{howe2006rise}. In the case of ER, one of the most difficult tasks is to decide whether two descriptions match or not. Crowd-sourced ER
assumes that humans can improve the effectiveness (i.e., accuracy) of Matching by leveraging contextual information and common sense. Therefore, it asks workers questions about the relation between descriptions for a small compensation per reply. Four main challenges arise in this context: 

~~~~~~~~~~~~~~Challenge 1: How should HITs be generated?
    
~~~~~~~~~~~~~~Challenge 2: How should HITs be formulated?

~~~~~~~~~~~~~~Challenge 3: How can we maximize accuracy, while minimizing the overall monetary cost?

~~~~~~~~~~~~~~Challenge 4: How can we restrict the labour cost?

\noindent
Below, we examine the main solutions to each challenge.

\vspace{3pt}
\noindent
\textbf{Challenge 1}: To generate HITs, a hybrid human-machine approach is typically used \cite{DBLP:conf/sigmod/LiZFWC17,DBLP:conf/gvd/Chen15}. First, machine-based techniques are used to do an initial, coarse pass over all pairs of candidate matches, discarding the majority of non-matches, and then, the crowd is asked to verify only the remaining candidate matches. This approach was first introduced by \textit{CrowdER} \cite{DBLP:journals/pvldb/WangKFF12}, which automatically computes the similarity between description pairs and discards those below a predetermined threshold. Similarly, \textit{ZenCrowd} \cite{DBLP:journals/vldb/DemartiniDC13} combines machine-based pre-processing with crowd-sourced matching, with the latter clarifying low confidence matches produced by the former. A probabilistic framework is used to refine crowd-sourced matches from inconsistent human responses.

\vspace{3pt}
\noindent
\textbf{Challenge 2}: Two are the main approaches to formulating HITs \cite{DBLP:conf/gvd/Chen15}: \textit{pair-based} and \textit{cluster-based} (a.k.a. \textit{multi-item}) \textit{HITs}. The former type asks workers questions of the form ``is $e_i$ matching with $e_j$?'' \cite{DBLP:journals/pvldb/VesdapuntBD14,DBLP:conf/sigmod/WangLKFF13,DBLP:journals/pvldb/WhangLG13,DBLP:journals/pvldb/FirmaniSS16,DBLP:conf/icde/VerroiosG15}, whereas the latter type involves groups with more than two descriptions, requesting workers to mark all duplicates within each group~\cite{DBLP:journals/pvldb/WangKFF12}. There is a trade-off between accuracy and efficiency in terms of cost and time between these two approaches~\cite{DBLP:conf/sigmod/VerroiosGP17}: pair-based HITs are simpler, allowing workers to provide more accurate responses,  
while the cluster-based HITs enable humans to mark many pairs of records with a few clicks, but their generation constitutes an NP-hard problem that is solved greedily by CrowdER~\cite{DBLP:journals/pvldb/WangKFF12}. 
\textit{Hybrid HITs} are used by \textit{Waldo} \cite{DBLP:conf/sigmod/VerroiosGP17}, which argues that the error rate of workers is different for different description pairs. Thus, the high error-rate pairs (i.e., the most ``difficult'' ones) should be formulated as pair-based HITs, whereas the low error-rate ones should form cluster-based HITs. Waldo formalizes the generation of the best hybrid HITs as an optimization task with a specific budget and provides solutions with probabilistic
guarantees. Finally, \textit{Crowdlink}~\cite{DBLP:conf/sigmod/ZhangM0Z15} decomposes each pair of descriptions into \textit{attribute-level HITs} to facilitate workers when processing descriptions with overwhelming information, i.e., with complex structures and attributes. A probabilistic framework then selects the $k$ best attributes.

\vspace{3pt}
\noindent
\textbf{Challenge 3}: To optimize the trade-off between accuracy and monetary cost, the transitive relation is typically leveraged; if the relation between two descriptions can be inferred by transitivity from the already detected duplicates, it is not crowd-sourced. This inference takes two flavours~\cite{DBLP:conf/gvd/Chen15}: \textit{positive transitivity} suggests that if $e_i \equiv e_j$ and $e_j \equiv e_k$, then $e_i \equiv e_k$, whereas \textit{negative transitivity} indicates that if $e_i \equiv e_j$, but $e_j \not\equiv e_k$, then $e_i \not\equiv e_k$. These relations lie at the core of several approaches~\cite{DBLP:journals/pvldb/WhangLG13,DBLP:conf/sigmod/WangLKFF13,DBLP:journals/pvldb/VesdapuntBD14,DBLP:journals/pvldb/FirmaniSS16,DBLP:journals/pvldb/KeTKY18} that minimize the number of HITs submitted to workers, reducing significantly the crowd-sourcing overhead. Their key insight is that finding matches before non-matches accelerates the ER process, by making the most of the transitive closure. 

Yet, these works assume that workers are infallible, operating as an oracle, which means that uncertainty comes exclusively from the machine-generated similarities. In practice, though, the high accuracy workers have an error rate up to 25\%, due to lack of domain expertise, individual biases, tiredness,  malicious behaviors as well as task complexity and ambiguity~\cite{DBLP:conf/cikm/YalavarthiKK17,DBLP:conf/sigmod/WangXL15}. When human errors occur, the above methods amplify them, thus compromising the overall ER accuracy~\cite{DBLP:conf/sigmod/WangXL15}. More realistic and robust approaches minimize HITs despite noisy workers, operating on top of a \textit{noisy matcher} that introduces uncertainty by returning possibly false results~\cite{DBLP:conf/cikm/YalavarthiKK17,DBLP:conf/icde/VerroiosG15,DBLP:conf/sigmod/ChaiLLDF16,DBLP:conf/cikm/KhanG16,DBLP:journals/vldb/ChaiLLDF18}. 
Other approaches correct the responses of an oracle through indirect ``control queries''~\cite{DBLP:conf/sigmod/GalhotraFSS18}, or refine the original crowd-sourced entities based on correlation clustering and additional HITs~\cite{DBLP:conf/sigmod/WangXL15}.

\vspace{3pt}
\noindent
\textbf{Challenge 4}: A major disadvantage of Crowd-sourced ER is the development cost that is required for applying it in practice. To address this issue, \textit{Corleone}~\cite{DBLP:conf/sigmod/GokhaleDDNRSZ14} implements a hands-off crowd-sourcing solution for the entire ER workflow that involves no software developers. It automatically generates blocking rules, learns a matcher from the HITs that are iteratively answered by workers (active learning minimizes the monetary cost), and finally returns the equivalence clusters. However, Corleone does not scale to large datasets, as it exclusively runs in-memory on a single machine. To address this issue, \textit{Falcon}~\cite{DBLP:conf/sigmod/DasCDNKDARP17} runs Corleone on a MapReduce cluster, exploiting crowd-time to run machine tasks. Experiments have shown that it scales to 2.5 million descriptions in 2-14 hours for only $\sim$\$60. \textit{CloudMatcher}~\cite{DBLP:journals/pvldb/GovindPNCDPFCCS18} goes one step further, implementing Falcon as a cloud service.

\vspace{-4pt}
\subsection{Rule-based ER methods}
\vspace{-2pt}

This category includes methods that leverage the knowledge of domain experts, who can provide some generic initial rules (e.g., ``if two descriptions have a similar address values, then they are matches'') that will help an ER algorithm to find some or all matches in a given task.

\textit{HIL}~\cite{DBLP:conf/edbt/HernandezKKPW13} is a high-level scripting language for expressing such rules. A HIL program determinescomplex~ER~pipelines, capturing the overall integration flow through a combination of SQL-like rules that link, map, fuse and aggregate descriptions. Its data model makes uses of logical indices to facilitate the modular construction and aggregation of complex entity descriptions. Its flexible, open type system allows HIL to handle irregular, sparse or partially known input data.

Reasoning and discovery techniques have also been proposed for automatically obtaining more matching rules.
Dependency-based reasoning techniques to help define keys for Matching and Blocking are introduced in \cite{DBLP:journals/pvldb/FanJLM09,DBLP:journals/vldb/FanGJLM11}. At their core lie \textit{matching dependencies} (\textbf{MDs}), which allow to infer matches, based on the similarity of database records on some attributes in relational schemata. MDs can be used in both Blocking and Matching to directly infer matches, but they can also be extended and used to infer new MDs, minimizing~manual~effort and leading to more matches.

Even though the MDs are looser versions of the strict functional dependencies in relational databases, they may still be too strict in practice. To address this issue, the \textit{conditional MDs} (\textbf{CMDs}) \cite{DBLP:journals/tkdd/WangSCYC17} bind MDs to a certain subset of descriptions in a relational table and have more expressive power than MDs for declaring constraints with conditions, allowing a wider range of~applications.

\textit{Certus}~\cite{DBLP:journals/pvldb/KwashieLLLSY19} introduces \textit{graph differential dependencies} (\textbf{GDDs}) as an extension of MDs and CMDs that enables approximate matching of values. It adopts a graph model for entity descriptions, which enables the formal representation of descriptions even in unstructured sources, while a specialized algorithm generates a non-redundant set of GDDs from labeled data. Certus employs the learned GDDs for improving the accuracy of ER results. Unlike MDs and CMDs, which operate only on structured data, Certus can identify matches irrespective of structure and with no assumed schema. 

\vspace{-4pt}
\subsection{Temporal ER methods}
\vspace{-2pt}

Entity descriptions are often associated with temporal information in the form of timestamps (e.g., user log data or sensor data) \cite{DBLP:journals/pvldb/ChiangDN14,DBLP:conf/pakdd/NanayakkaraCR19} or temporal validity of attributes (e.g., population, marital status, affiliation) \cite{DBLP:journals/ai/HoffartSBW13}. ER methods exploiting such temporal information may show better performance than those ignoring it~\cite{DBLP:conf/sigmod/ChiangDN14}; rather than deciding if two descriptions match, they try to decide if a new description matches with a set descriptions that have been already identified as matches. The probability of a value re-appearing over time is examined in  \cite{DBLP:conf/sigmod/ChiangDN14}. Intuitively, a description might change its attribute values in a way that is dependent on previous values. For example, if a person's location has taken the values Los Angeles, San Francisco, San Jose in the past, then these values are more likely to appear in this person's future location than Berlin or Cairo. \textit{SFDS} \cite{DBLP:journals/pvldb/ChiangDN14} follows a ``static first, dynamic second'' strategy: initially, it assumes that all descriptions are static (i.e., not evolving over time) and groups them into clusters. These are later merged in the dynamic phase, if the different clusters correspond to the same entities that have evolved~over~time. 

\subsection{Open-source ER tools}
\label{sec:tools}

\begin{table*}[t!]\centering
\vspace{-5pt}
\caption{The main open-source ER Tools (a feature in parenthesis is partially supported).}
\vspace{-5pt}
\label{tab:erTools}
\setlength{\tabcolsep}{2.7pt}
{\scriptsize
\begin{tabular}{|l|c|c|c|c|c|c|c|c|c|}
\hline
\textbf{Tool} & \textbf{Blocking} & \textbf{Block} & \textbf{Matching} & \textbf{Clustering} & \textbf{Parallelization} & \textbf{Bugdet-} & \textbf{Incremental} & \textbf{GUI} & \textbf{Language} \\ 
& & \textbf{Processing} & & & & \textbf{aware ER} & \textbf{ER} & & \\
\hline
\hline
Dedupe \cite{BilenkoM03} & \checkmark & - & \checkmark & - & multi-core & - & - & - & Python	\\
\hline
DuDe \cite{draisbach2010dude} & \checkmark & - & \checkmark & - & - & -	& - & - & Java \\
\hline
Febrl \cite{Christen08}	& \checkmark & - & \checkmark & - & multi-core	& - & - & \checkmark & Python \\
\hline
FRIL \cite{jurczyk2008fine}	& \checkmark & - & \checkmark & - & - & - & - & \checkmark	&  Java	\\
\hline
OYSTER \cite{nelson2011entity}	& \checkmark & - & \checkmark & - & - &	- & - & - & Java	\\	
\hline
RecordLinkage \cite{sariyar2011controlling}	& \checkmark & - & \checkmark & - & - & - & - & - & R\\
\hline
Magellan \cite{DBLP:journals/pvldb/KondaDCDABLPZNP16} &	\checkmark & - & \checkmark & - & (Apache Spark) & - & - & \checkmark & Python\\	
\hline
FAMER \cite{DBLP:journals/csimq/SaeediNPR18} & - & - & - & \checkmark & Apache Flink & - & - & - & Java \\
\hline
\hline
Silk \cite{DBLP:journals/pvldb/IseleB12}	&	\checkmark & - & \checkmark & - & Apache Spark & - & - & \checkmark &	Scala	\\
\hline
LIMES \cite{NgomoA11} & \checkmark & - & \checkmark & - & (multi-core) & - & - & \checkmark & Java	\\
\hline
Duke	&	\checkmark	& - & \checkmark & - & - & - & \checkmark & - & Java	\\
\hline
KnoFuss  \cite{DBLP:conf/ekaw/NikolovUMR08}	&	\checkmark	& - & \checkmark & - & - & - & - & - & Java	\\
\hline
SERIMI  \cite{DBLP:journals/tkde/AraujoTVS15}	&	\checkmark	& - & \checkmark & - & - & - & - &	- & Ruby	\\
\hline
\hline
MinoanER \cite{DBLP:conf/edbt/Efthymiou0SC19} &	\checkmark	& \checkmark & \checkmark	&	- & Apache Spark & - &	- & - & Java	\\
\hline
JedAI \cite{DBLP:journals/pvldb/PapadakisTTGPK18}	&	\checkmark	& \checkmark & \checkmark	& \checkmark & Apache Spark & \checkmark & - & \checkmark & Java\\
\hline
\end{tabular}
}
\vspace{-5pt}
\end{table*}

We now elaborate on the main systems that are crafted for end-to-end Entity Resolution. We examined the 18 non-commercial and 15 commercial tools that are listed in the extended version of \cite{DBLP:journals/pvldb/KondaDCDABLPZNP16}\footnote{{\scriptsize\url{http://pages.cs.wisc.edu/~anhai/papers/magellan-tr.pdf}}} along with the 10 Link Discovery frameworks surveyed in \cite{DBLP:journals/semweb/NentwigHNR17}. Among them,
we exclusively consider the open-source systems, since the closed-code and the commercial ones provide insufficient information about their internal functionality and/or their algorithms.

A summary of the main open-source ER systems appears in Table \ref{tab:erTools}. For each one, we report whether it involves one or more methods per workflow step of the general end-to-end ER pipeline in Figure \ref{fig:workflow}(a), whether it supports parallelization, budget-aware or incremental methods, a graphical user interface (GUI) as well as its programming language. To facilitate their understanding, we group all systems into 3 categories, depending on their input data: (i) systems for structured data, (ii) systems for semi-structured data, and (iii) hybrid systems.

The tools for structured data include Dedupe \cite{BilenkoM03}, FRIL \cite{jurczyk2008fine}, OYSTER \cite{nelson2011entity}, RecordLinkage \cite{sariyar2011controlling}, DuDe \cite{draisbach2010dude}, Febrl \cite{Christen08}, Magellan \cite{DBLP:journals/pvldb/KondaDCDABLPZNP16} and FAMER \cite{DBLP:journals/csimq/SaeediNPR18}. All of them offer at least one method for Blocking and Matching, while disregarding Clustering. The only exception is FAMER, which exclusively focuses on Clustering, implementing several established techniques in Apache Flink. Febrl involves the richest variety of non-learning, schema-aware Blocking methods, which can be combined with several similarity measures and top-performing classifiers for supervised matching. Magellan conveys a Deep Learning module, which is a unique feature among all ER tools. Most systems are implemented in Java or Python, with just three of them offering a GUI.

The systems for semi-structured data receive as input RDF dump files or SPARQL endpoints. The most prominent ones are Silk \cite{DBLP:journals/pvldb/IseleB12} and LIMES \cite{NgomoA11}, which are crafted for the Link Discovery problem (i.e., the generic task of identifying relations between entities). Restricting them to the discovery of \texttt{sameAs} relations renders them suitable for ER. Both systems involve custom blocking techniques along with a large variety of character- and token-based similarity measures. Combinations of these similarity measures are learned in a (semi-)supervised way for effective Matching. Both tools offer an intuitive GUI, unlike the remaining ones, namely SERIMI \cite{DBLP:journals/tkde/AraujoTVS15}, Duke\footnote{\label{dukeURL}\url{https://github.com/larsga/Duke}} and KnoFuss \cite{DBLP:conf/ekaw/NikolovUMR08}. These systems merely apply simple Blocking techniques to literal values and focus primarily on Matching, providing effective, but custom techniques based on similarity measures. 

The hybrid tools, MinoanER \cite{DBLP:conf/edbt/Efthymiou0SC19} and JedAI \cite{DBLP:journals/pvldb/PapadakisTTGPK18}, apply uniformly to both structured and semi-structured data. This is possible due to the schema-agnostic functionality of their methods. In fact, they implement the main non-learning, schema-agnostic techniques for Blocking, Matching and Clustering. They are also the only systems that offer Block Processing techniques. 

Overall, we observe that all open-source systems focus on Matching, conveying a series of string similarity measures for the comparison of attribute values. More effort should be spend on covering more adequately all workflow steps of the general end-to-end ER workflow. Most importantly, except for Duke's Incremental ER and JedAI's Progressive ER, no system supports any other processing mode other than budget-agnostic ER. This should be addressed in the future.

\subsection{Discussion}

Even though Rule-based and Temporal ER constitute important topics, more effort is lately directed at leveraging Deep Learning techniques for ER. These efforts have already paid off, as the resulting techniques achieve the state-of-the-art performance for several established benchark datasets \cite{DBLP:conf/sigmod/MudgalLRDPKDAR18}, outperforming methods based on traditional machine learning. Yet, the time efficiency and the availability of  a representative set of labelled instances remain important issues. The latter is intelligently addressed by a series of Crowdsourcing-based ER methods. Despite the considerable recent advancements, though, Crowdsourced ER still suffers from significant monetary cost and high latency, while it can only be used by expert users. Systems like CloudMatcher contribute to its democratization, while systems like MinoanER and JedAI aim to act as libraries of the state-of-the-art methods for end-to-end ER over Big Data.

\section{Directions for future work} 
\label{sec:directions}
\vspace{-2pt}

As we have just begun to realize the need for \emph{Entity Resolution Management Systems}~\cite{DBLP:journals/pvldb/KondaDCDABLPZNP16}, we next highlight few critical research directions for future work, which aim to support advanced services for specifying,  maintaining and making accountable complex ER workflows.

\vspace{2pt}
\noindent \textbf{Multi-modal ER.} In the Big Data era, multi-modal entity descriptions are becoming increasingly common. 
Factual, textual or image-based descriptions of the same real world entities are available from different sources and at different temporal, or spatial resolutions. 
Each modality carries a specific piece of information about an entity and offers added value that cannot be obtained from the other modalities. 
Recent years have witnessed a surge of the need to jointly analyze multi-modal descriptions~\cite{Zheng2018MultiModalCR}. 
Finding semantically similar descriptions from different modalities is one of the core problems of multi-modal learning. 
Most current approaches focus on how to utilize extrinsic supervised information to project one modality to the other, or map two modalities into a commonly shared space. 
The performance of these methods heavily depends on the richness of the training samples. 
In real-world applications though, obtaining matched data from multiple modalities is costly, or impossible~\cite{Gao:2018:CMS:3169405.3169727}. 
Thus, we need sample-insensitive methods for multi-modal ER, and in this respect, we can leverage recent advances in multi-modal ML techniques~\cite{Baltrusaitis:2018:CAM:3107990.3107993}.

\vspace{2pt}
\noindent \textbf{Debugging and Repairing ER workflows.} 
Current ER research mainly focuses on developing accurate and efficient techniques, which in reality are constrained by a number of factors, such as low quality entity descriptions, ambiguous domain knowledge and limited ground truth. 
Hence, it is difficult to guarantee the quality of ER workflows at specification time. 
To support a \emph{continuous} specification of ER workflows, an iterative approach is needed to refine ER workflows by identifying and analyzing the mistakes (false matches and non-matches) of ER enactments at each iteration step. 
Debugging ER workflows requires to: (a) understand the mistakes made by Blocking or Matching algorithms; (b) diagnose root-causes of these mistakes (e.g., due to dirty data, problematic feature sets, or even tuning parameters of algorithms); and (c) prioritize mistakes and take actions to fix them~\cite{DBLP:journals/pvldb/KondaDCDABLPZNP16}. 
We note that not all categories of mistakes have the same impact on the end-to-end quality of ER workflows. 
For example, the removal of outliers from input data often leads to overfitting problems of learning-based matchers. 
Recognizing patterns of mistakes reproduced under similar conditions can provide valuable insights in order to repair ER workflows. 
The focus of ER work so far was in preventing rather than repairing mistakes in ER results. 
Recent work on debugging and repairing Big Data analytics pipelines can be leveraged in this respect~\cite{Chung19,Logothetis:2013:SLC:2523616.2523619,Gulzar16}.

\vspace{2pt}
\noindent \textbf{Fairness in Long Tail Entities Resolution.} The reported accuracy scores of several ER approaches are fairly high, giving the impression that the problem is well-understood and solved. At the same time, recent works (e.g.,~\cite{DBLP:conf/lrec/ErpMPIPRW16,DBLP:conf/ecir/EsquivelAMCM17}) claim that ER systems base their performance on entity popularity, while their performance drops significantly when focusing on the rare, long tail entities. However, the lack of formal definitions regarding what is popular and long tail entities for the ER task prevents the identification of the difficult cases for ER, for which systems need to be adapted or new approaches need to be developed~\cite{DBLP:journals/debu/WangHM18}. Better understanding such cases will be helpful for ER, since knowledge about long tail entities is less accessible, not redundant and hard to obtain. 

\vspace{2pt}
\noindent\textbf{Diversity of Matching Entities.} Works in budget-aware ER typically focus on maximizing the reported matches, by potentially exploiting the partial matching results obtained so far in an iterative process. 
Then, it will be interesting to measure the added knowledge that the ER process could achieve after merging the matches, similar to the notion of diversity in information retrieval. Our intuition is that merges resulting from somehow similar entities are more beneficial when compared to merges from strongly similar entities. Thus, given a constraint in the number of possible merges, the goal is to perform those that contribute most in diversifying the knowledge encoded in the result. Added knowledge can be measured by the number of relationships of a merged entity with other entities. We consider such relationships as a unit of knowledge increase: when two relationships represent two different knowledge units, they are both useful; when they overlap, they represent the same knowledge unit, so we do not gain by knowing both of them. 

\vspace{2pt}
\noindent\textbf{Bias in ER.} Similarity measures lie at the core of Matching.
However, it is well known that not all similarity measures are appropriate for all types of data (e.g., strings, locations, and videos). Moreover, when focusing on particular types of measures, e.g., measures for string matching, we do not know beforehand which is the ideal measure for counting similarities with respect to the semantics of the strings to be compared. For instance, we possibly need different measures for computing similarities between American names than for Chinese names. In such scenarios, we typically exploit some solid empirical evidence, which, based on some of the characteristics that our data have, leads us to select, unintentionally, a particular measure. This fact can be considered as algorithmic bias \cite{DBLP:conf/kdd/HajianBC16}. As a first step, for achieving unbiased and fair results, it is important to experimentally study if there is bias in ER algorithms \cite{DBLP:conf/edbt/0001P19,DBLP:conf/sigsoft/AngellJBM18}. Moving forward to the next generation of approaches, we need to propose solutions and provide guidelines that make ER algorithms fair. 

\section{Conclusions}
\label{sec:conclusions}
Although ER has been studied for more than three decades in different computer science communities, it still remains an active area of research. The problem has enjoyed a renaissance during recent years, with the avalanche of data-intensive descriptions of real-world entities provided by government, scientific, corporate or even user-crafted data sources. Reconciling different entity descriptions in the Big Data era poses new challenges both at the algorithmic and the system level: Volume, due to the very high number of entities and data sources, Variety, due to the extreme schema heterogeneity, Velocity, due to the continuously increasing volume of data, and Veracity, due to the high level of noise and inconsistencies. In this survey, we have focused on how the main algorithms in each step of the end-to-end ER workflow address the combination of these challenges. 
Blocking and Block Processing, two steps that by definition tackle Volume, also address Variety mainly through a schema-agnostic, non-learning functionality. Most Matching methods employ a schema-agnostic, collective functionality, which leverages information provided by related entities, in order to address Variety and Veracity. Budget-aware ER methods rely on Blocking and a usually schema-agnostic functionality to simultaneously address Volume and Variety, while Incremental Methods address Volume and Velocity through Blocking, but their schema-aware functionality prevents them from tackling Variety, too. In all cases, massive parallelization, usually through the MapReduce framework, plays an important role in further improving scalability and, thus, addressing Volume. Note, though, that we share the view of ER as an engineering task by nature, and hence, we cannot just keep developing ER algorithms in a vacuum~\cite{DBLP:journals/pvldb/KondaDCDABLPZNP16}. In the Big Data era, we opt for \emph{open-world ER systems} that allow to plug-and-play different algorithms and can easily integrate with third-party tools for data exploration, data cleaning or data analytics. 

\vspace{-1pt}
\def\thebibliography#1{

\noindent
\textbf{Acknowledgements.} This work was partially funded by the EU H2020 project ExtremeEarth (825258).

\section*{References}
	\scriptsize
  \list
    {[\arabic{enumi}]}
    {\settowidth\labelwidth{[#1]}
     \leftmargin\labelwidth
     \parsep 0pt                
     \itemsep 0pt               
     \advance\leftmargin\labelsep
     \usecounter{enumi}
    }
  \def\newblock{\hskip .11em plus .33em minus .07em}
  \sloppy\clubpenalty10000\widowpenalty10000
  \sfcode`\.=1000\relax
}


\bibliographystyle{abbrv}
\bibliography{Dissertation}
\end{document}